\documentclass{aa}  

\usepackage[colorlinks=true,allcolors=blue]{hyperref}
\usepackage{threeparttable}
\usepackage{subcaption}

\usepackage{graphicx}
\usepackage{txfonts}
\usepackage{xspace}
\newcommand{\kms}{km~s$^{-1}$\xspace}
\newcommand{\ergs}{erg~s$^{-1}$\xspace}
\newcommand{\Msunyr}{M$_{\odot}$ yr$^{-1}$\xspace}

\newcommand{\hb}{H$\beta$\xspace}
\newcommand{\ha}{H$\alpha$\xspace}
\newcommand{\lya}{Ly$\alpha$\xspace}
\newcommand{\heii}{He\,{\sc{ii}}\xspace}
\newcommand{\feii}{Fe\,{\sc{ii}}\xspace}

\newcommand{\oiiiuv}{O\,{\sc{iii}}]\xspace}

\newcommand{\oiii}{[O\,{\sc{iii}}]\xspace}

\newcommand{\sii}{[S\,{\sc{ii}}]\xspace}

\newcommand{\civ}{C\,{\sc{iv}}\xspace}
\newcommand{\ciii}{C\,{\sc{iii}}]\xspace}

\newcommand{\nv}{N\,{\sc{v}}\xspace}

\newcommand{\fevii}{[Fe\,{\sc{vii}}]\xspace}
\newcommand{\nev}{[Ne\,{\sc{v}}]\xspace}

\newcommand{\nii}{[N\,{\sc{ii}]}\xspace}

\newcommand{\chandra}{\textit{Chandra}\xspace}

\begin{document}

   \title{GA-NIFS: High number of dual active galactic nuclei at $z\sim 3$}

   %\subtitle{I. Overviewing the $\kappa$-mechanism}

   \author{Michele Perna
          \inst{\ref{iCAB}}\thanks{e-mail: mperna@cab.inta-csic.es}
          \and
          Santiago Arribas\inst{\ref{iCAB}}
          \and 
          Isabella Lamperti\inst{\ref{iCAB},\ref{iUNIFI}, \ref{iOAA}}
          \and
          Chiara Circosta\inst{\ref{iESAes}}
          \and
          Elena Bertola\inst{\ref{iOAA}}
          \and
          Pablo~G.~P\'erez-Gonz\'alez\inst{\ref{iCAB}}
          \and
          Francesco D'Eugenio\inst{\ref{iKav},\ref{iCav}}
          \and
          Hannah \"{U}bler\inst{\ref{iKav},\ref{iCav}, \ref{iMPE}}
          \and
          Giovanni Cresci\inst{\ref{iOAA}}
          \and
          Marta Volonteri\inst{\ref{iIAP}}
          \and
          Filippo~Mannucci\inst{\ref{iOAA}}
          \and
          Roberto~Maiolino\inst{\ref{iKav},\ref{iCav},\ref{iUCL}}
          \and
          Bruno Rodr\'iguez~Del~Pino\inst{\ref{iCAB}}
          \and
          Torsten Böker\inst{\ref{iESAusa}}
          \and
          Andrew~J.~Bunker\inst{\ref{iOxf}}
          \and
          St\'ephane~Charlot\inst{\ref{iSor}}
          \and
          Chris~J.~Willott\inst{\ref{iNRC}}
          \and
          Stefano~Carniani\inst{\ref{iNorm}}
          \and
          Mirko Curti\inst{\ref{iESOge}}
          \and
          Gareth C.~Jones\inst{\ref{iOxf},\ref{iKav},\ref{iCav}}
          \and
          Nimisha~Kumari\inst{\ref{iESAusa}}
          \and
          Madeline~A.~Marshall\inst{\ref{iLOS}}
          \and
          Giacomo~Venturi\inst{\ref{iNorm}}
          \and
          Aayush~Saxena\inst{\ref{iOxf}}
          \and 
          Jan Scholtz\inst{\ref{iKav},\ref{iCav}}
          \and 
          Joris Witstok\inst{\ref{iKav},\ref{iCav}}
          }

   \authorrunning{M. Perna et al.}
   \institute{
            Centro de Astrobiolog\'ia (CAB), CSIC--INTA, Cra. de Ajalvir Km.~4, 28850 -- Torrej\'on de Ardoz, Madrid, Spain\label{iCAB}
    \and
            Università di Firenze, Dipartimento di Fisica e Astronomia, via G. Sansone 1, 50019 Sesto F.no, Firenze, Italy\label{iUNIFI}
    \and
            INAF - Osservatorio Astrofisico di Arcetri, Largo E. Fermi 5, I-50125 Firenze, Italy\label{iOAA}
    \and
            Kavli Institute for Cosmology, University of Cambridge, Madingley Road, Cambridge, CB3 0HA, UK\label{iKav}
    \and
            Cavendish Laboratory - Astrophysics Group, University of Cambridge, 19 JJ Thomson Avenue, Cambridge, CB3 0HE, UK\label{iCav}
    \and
            Max-Planck-Institut f\"ur extraterrestrische Physik (MPE), Gie{\ss}enbachstra{\ss}e 1, 85748 Garching, Germany\label{iMPE}
    \and
            Institut d’Astrophysique de Paris, UPMC et CNRS, UMR 7095, 98 bis bd Arago, F-75014 Paris, France\label{iIAP}
    \and
            Department of Physics and Astronomy, University College London, Gower Street, London WC1E 6BT, UK\label{iUCL}  
    \and
            Department of Physics, University of Oxford, Denys Wilkinson Building, Keble Road, Oxford OX1 3RH, UK\label{iOxf}
    \and
            Sorbonne Universit\'e, CNRS, UMR 7095, Institut d’Astrophysique de Paris, 98 bis bd Arago, 75014 Paris, France\label{iSor} 
    \and
            Scuola Normale Superiore, Piazza dei Cavalieri 7, I-56126 Pisa, Italy\label{iNorm}
    \and
            European Southern Observatory, Karl-Schwarzschild-Strasse 2, 85748 Garching, Germany\label{iESOge}
    \and    
            European Space Agency, c/o STScI, 3700 San Martin Drive, Baltimore, MD 21218, USA\label{iESAusa}
    \and    
            European Space Agency, ESAC, Villanueva de la Ca\~{n}ada, E-28692 Madrid, Spain\label{iESAes}
    \and
            NRC Herzberg, 5071 West Saanich Rd, Victoria, BC V9E 2E7, Canada\label{iNRC}
    \and
            Los Alamos National Laboratory, Los Alamos, NM 87545, USA\label{iLOS}
             }

   \date{Received September 15, 1996; accepted March 16, 1997}

% \abstract{}{}{}{}{} 
% 5 {} token are mandatory

  \abstract
   {Merger events can trigger gas accretion onto supermassive black holes (SMBHs) located at the centre of galaxies and form close pairs of active galactic nuclei (AGNs). The fraction of AGNs in pairs offers critical insights into the dynamics of galaxy interactions, SMBH growth, and their co-evolution with host galaxies. 
   However, the identification of dual AGNs is difficult, as it requires high-quality spatial and spectral data; hence,  very few pairs have been found in the distant Universe so far.
   }
   {This study is aimed at providing a first observational estimate of the fraction of dual AGNs at $2 < z < 6$ by analysing a sample of 16 AGNs observed with the JWST Near-InfraRed Spectrograph (NIRSpec) in integral field mode, as part of the GA-NIFS survey. For two AGNs in our sample, we also incorporated archival VLT/MUSE data to expand the search area.}
   {We searched for nearby companion galaxies and emission-line sources within the $\sim 20\times 20$~kpc field of view of the NIRSpec data cubes, extending up to $\sim50$~kpc using the MUSE data cubes. We analysed the spectra of such emitters to determine their physical and kinematic properties.}
   {We report the serendipitous discovery of a triple AGN system and four dual AGNs (two of which had been considered as candidates), with projected separations in the range 3--28 kpc. The results of this study more than double the number of known multiple AGNs at $z > 3$ at these separations. Their AGN classification is mainly based on standard optical emission line flux ratios, as observed with JWST/NIRSpec, and complemented with additional multi-wavelength diagnostics. The identification of these 3-5 multiple AGNs out of the 16 AGN systems in the GA-NIFS survey (i.e. $\sim$ 20--30\%) suggests they might be more common than previously thought from other observational campaigns. Moreover, our inferred fraction of dual AGN moderately exceeds predictions from cosmological simulations that mimic our observational criteria ($\sim 10$\%).}
   {This work highlights the exceptional capabilities of NIRSpec for detecting distant dual AGNs, prompting new investigations to constrain their fraction across cosmic time.}


   \maketitle
%
%-------------------------------------------------------------------

\section{Introduction}

Since supermassive black holes (SMBHs) are believed to be present at the centre of most massive galaxies (e.g. \citealt{Magorrian1998}), merging systems with two or more SMBHs are a natural and expected occurrence (e.g. \citealt{Koss2012, Koss2023}).  Studies of these objects have been receiving increasing attention, as they can be used to constrain the link between galaxy mergers and SMBH feeding and feedback effects across cosmic time, which is a key ingredient for galaxy evolution models (\citealt{DeRosa2019,Volonteri2022,Chen2023}). 
Moreover, these SMBH pairs are the parent population of the systems that will eventually merge, producing the gravitational wave signals now detected by the pulsar timing array (PTA) experiments (e.g. \citealt{Antoniadis2023,Agazie2023a}) for SMBHs with masses in the range 10$^{8}$--10$^9$~M$_\odot$. Similarly, mergers involving lighter SMBHs (10$^{4}$--10$^7$~M$_\odot$) are expected to be detectable with the upcoming Laser Interferometric Space Antenna (LISA; e.g. \citealt{Agazie2023b}).
However, beyond the local Universe (e.g. \citealt{Voggel2022}), systems with multiple SMBHs can only be detected when both nuclei shine as active galactic nuclei (AGNs) simultaneously. This phenomenon occurs during brief active phases within the much longer galaxy merging process (e.g. \citealt{Capelo2017}).

Cosmological simulations predict a fraction of AGN pairs, separated by hundreds to few thousands parsecs, out of the total number of AGNs, of the order of a few percent in the redshift range $1<z<5$, depending  on the selection criteria on luminosity, distance, and mass of the selected systems (e.g. \citealt{DeRosa2019,Volonteri2022,Chen2023, PuertoSanchez2024}). Observations have been consistent to date with these theoretical predictions; so far, only very few dual AGNs have been found at $z>1$ (e.g.  \citealt{Lemon2023, Mannucci2022, Mannucci2023}).

The detection of dual AGNs is hampered by several challenges. The spatial resolution and sensitivity constraints often preclude resolving small separations, especially in distant galaxies. Moreover, the distinction between dual AGNs and other phenomena, such as gravitational lensing effects, requires high-quality spectral and spatial data, often requiring dedicated follow-up observations. 
Many dual AGNs have been discovered serendipitously (e.g. \citealt{Shields2012, Husemann2018a, Izumi2024}) thanks to these limitations (\citealt{DeRosa2019}). 
However, more recently, systematic searches have been initiated thanks to the development of new techniques taking full advantage of the all-sky Gaia database (\citealt{Gaia2018}). 
The Gaia-Multi-Peak (GMP, \citealt{Mannucci2022}) technique facilitates the identification of dual AGN candidates at separations in the range of 0.1-0.7\arcsec\ (e.g. 0.8--5.5 kpc at $z=3$), searching for multiple peaks in the light profile of the Gaia sources. Another method is the Varstrometry technique (e.g. \citealt{Hwang2020}), which extends systematic searches for dual AGN candidates down to even lower separations by exploiting the temporal displacements of the  photocentre of variable, unresolved sources. 
These new methods have been shown to be effective in discovering new dual AGN candidates (up to $z\sim 3$; \citealt{Mannucci2023, Lemon2023}). However, they are generally limited to the relatively bright targets detected by Gaia, requiring the two AGN components to have comparable luminosities, and needing dedicated spectroscopic follow-up strategies to spatially resolve the candidate AGN pairs. Moreover, spectroscopic observations are required to distinguish multiple spatial components due to gravitational lensing of a single active galaxy from real AGN pairs (\citealt{Lemon2023, Scialpi2024}). 
These methods also have an important observational bias, as they require these systems to have significant emission in the optical band ($\sim 4000-9500$~\AA)  to be selected with Gaia. As a consequence, all dual AGNs discovered with Gaia are unlikely to be heavily affected by dust extinction. This is an important limitation, since an enhanced level of AGN obscuration is actually expected in late-stage mergers (e.g. \citealt{Hopkins2016, Ricci2017, Perna2018, Koss2018}). Therefore, the dual AGNs confirmed so far only provide a lower limit to their actual prevalence, especially at high redshifts (e.g. \citealt{Li2024}).

With its orders of magnitude improvement in sensitivity and resolution across the wavelength range from 0.6 to 29 $\mu$m compared to the previous generation of infrared telescopes (\citealt{Rigby2023}), JWST reveals our Universe in a whole new light. Specifically, it is opening the opportunity to study in detail the environment of AGNs in the early Universe by probing their emission in the rest-frame ultraviolet (UV) and optical spectral ranges. The first JWST spatially resolved spectroscopic observations, for instance obtained with the integral field spectrograph (IFS) of NIRSpec (\citealt{Jakobsen2022, Boker2022}) and the wide-field slitless spectroscopy (WFSS) mode of NIRCam (\citealt{Greene2017}) have already revealed that high-$z$ AGNs are often surrounded by close companions (\citealt{Kashino2022, Wylezalek2022, Yang2023,Perna2023, Marshall2023, Marshall2024,Ubler2023,Ubler2024zs7, Schindler2024}). 
Moreover, JWST observations show tidal bridges and tails at kiloparsec scales connecting the AGN with the companions, thereby revealing the presence of gravitational interactions up to $z\sim 7$ (e.g. \citealt{Marshall2023, Loiacono2024}). These spectroscopic data also help unveil the nature of the AGN companions; for instance, classical emission line flux ratio diagnostics such as the optical and UV diagrams (e.g. \citealt{Baldwin1981, Feltre2016, Nakajima2022}) can be used to identify multiple AGNs (e.g. \citealt{Perna2023, Zamora2024}).

%GA-NIFS
In this work, we report the discovery of an  AGN triplet, along with two secure and two candidate dual AGNs at $z\sim 3$. They have projected separations between 3 and 28~kpc and are located in the GOODS-S (\citealt{Giavalisco2004}) and COSMOS fields (\citealt{Scoville2007}). Importantly, this work doubles the number of known multiple AGN systems at $z\gtrsim 3$ with separations $<30$~kpc (\citealt{Mannucci2023}). All AGN pairs have been observed with NIRSpec IFS; the triple AGN has been instead identified by combining NIRSpec and archival observations from the optical IFS Multi Unit Spectroscopic Explorer (MUSE) at the Very Large Telescope (VLT). 
The sample of multiple AGN systems was drawn from a larger sample of 16 AGN systems at $2<z<6$, targeted as part of the JWST programme ``Galaxy Assembly with NIRSpec IFS'' (GA-NIFS\footnote{\url{https://ga-nifs.github.io/}}). Some of these 16 targets have been already presented in the literature by our team and they appear in \citet[the target GS539, also know as Aless073.1]{Parlanti2024aless}, \citet[GS3073]{Ubler2023}, \citet[Jekyll \& Hyde]{Perez-Gonzalez2024}, \citet[GS10578]{DEugenio2024NatAs}, and \citet[GS133]{Perna2024gs133}.

The paper is outlined as follows. Section \ref{sec:sample} introduces the GA-NIFS survey and the sample of AGNs analysed in this work.
In Sect. \ref{sec:observations} we describe our JWST/NIRSpec IFS observations and data reduction procedures, as well as the archival data used to complement our analysis. The detailed data analysis of spectroscopic data is reported in Sect. \ref{sec:analysis}. 
The optical spectra of our dual AGNs and emission line diagnostics employed for their classification are presented in Sect. \ref{sec:resultsOPT}. The UV spectra and diagnostics used to classify the third active nucleus in one of our systems are detailed in Sect. \ref{sec:resultsUV}. 
The AGN and host galaxy properties of these systems are examined in Sect. \ref{sec:bolometricluminosities}-\ref{sec:xray}. 
Finally, we discuss our results and present our estimate of the dual AGN fraction in Sect. \ref{sec:discussion}. We conclude with a summary of our findings in Sect. \ref{sec:conclusions}.
Throughout, we adopt a \cite{Chabrier2003} initial mass function ($0.1-100~M_\odot$) and a flat $\Lambda$CDM cosmology with $H_0=70$~km~s$^{-1}$~Mpc$^{-1}$, $\Omega_\Lambda=0.7$, and $\Omega_m=0.3$. 
In our analysis of NIRSpec data, we use vacuum wavelengths according to their calibration. 
Similarly, for MUSE data, we use vacuum wavelengths, consistent with the standard pre-JWST practice.

\section{GA-NIFS sample of AGNs at $2<z<6$}\label{sec:sample}

GA-NIFS is a JWST Guaranteed Time Observations (GTO) programme  aimed at characterising the internal structure and close environment of a sample of $z>2$ galaxies, as well as to investigate the physical processes driving galaxy evolution across cosmic time. It plans to observe 55 targets with NIRSpec IFS in JWST cycles 1 and 3, including both star-forming galaxies (SFGs) and AGNs (e.g. \citealt{Jones2024HFLS3, Arribas2024, Parlanti2024hz4, RodriguezDelPino2024, Ubler2024gn20}). 

The GA-NIFS sub-sample of non-active galaxies were heterogeneously selected among the most luminous and/or extended sources (as observed in Ly$\alpha$, NIR, and/or mm wavelengths), together with a few well studied sources (e.g. Jekyll \& Hyde; \citealt{SchreiberC2018,Perez-Gonzalez2024}) to increase the leverage of IFS capabilities. 

The sub-sample of active galaxies includes sources at $z<6$ residing in the COSMOS and GOODS-S fields and selected solely for their X-ray emission $L_{2-10 \ {\rm keV}} \gtrsim 10^{44}$ \ergs (from \citealt{Marchesi2016} and \citealt{Luo2017}), which is unambiguously attributed to AGN emission. This sub-sample consists of 7 sources in GOODS-S  
and 6 in COSMOS.  
A careful inspection of the NIRSpec data of other GA-NIFS targets in the same cosmological fields -- but not fitting in the above X-ray selection criterion -- allowed us to identify additional targets hosting AGN: we unveiled the presence of AGN activity, based on BPT diagnostics (\citealt{Baldwin1981}), in two sources originally selected as non-active galaxies, GS19293 (Venturi et al., in prep.), and in the close surroundings of Jekyll (\citealt{Perez-Gonzalez2024}).  
GS3073 must also be considered an AGN, due to the detection of broad line region (BLR) emission in its rest-frame optical spectrum (\citealt{Ubler2023}).

We note that the GA-NIFS survey also includes  additional AGNs that have not been considered in this study. They were specifically selected due to their dense environments: LBQS 0302$-$0019 at $z\sim 3.3$, with its own secondary obscured AGNs, presented in \cite{Perna2023}; BR1202$-$0725 at $z\sim 4.7$, also featuring a secondary AGN identified through NIRSpec IFS data, presented in \cite{Zamora2024}. Since they reside in well-known overdensities, these two systems were excluded from this work, as their inclusion would bias our search of dual AGNs. We also excluded GS5001, a source we presented in \citet{Lamperti2024}. Although GS5001 is detected in X-rays, it remains uncertain whether it hosts an AGN (see also \citealt{Lyu2024}). Additionally, since GS5001 is situated in a well-known overdense region, it was excluded from this work to maintain consistency in our sample selection. Similarly, we excluded the GN20 source we presented in \citet{Ubler2024gn20}, which also resides in an overdensity (e.g. \citealt{Daddi2009, CrespoGomez2024}).

Additional targets at $z>6$ from GA-NIFS (e.g. \citealt{Christensen2023, Marshall2023}) were not taken into account in this work because in these high-$z$ sources the \nii$\lambda6583$ is very faint and usually undetected. This is likely due to their lower metallicity (e.g. \citealt{Kocevski2023, Ubler2023, Maiolino2023c, Ji2024gs3073}), or because the line is redshifted out of the spectral range of NIRSpec (e.g. \citealt{Marshall2024}); hence, it precludes the use of BPT diagnostics to discover AGN emission in their companion sources. Other diagnostics might be required for these systems (e.g. the detection of faint \oiii$\lambda4363$,  \heii$\lambda$4686 and BLR lines in the optical spectra; see e.g. \citealt{Nakajima2022}, \citealt{Tozzi2023}, \citealt{Ubler2024zs7}, and \citealt{Mazzolari2024}) and we therefore defer this analysis to a forthcoming paper.

\begin{figure*}[!h]
\centering
\includegraphics[width=0.23\textwidth, trim=0mm 0mm 0mm 4mm]{{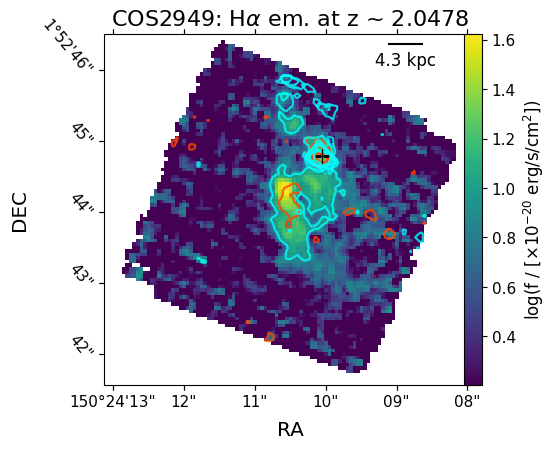}}
\includegraphics[width=0.23\textwidth, trim=0mm 4mm 0mm 4mm]{{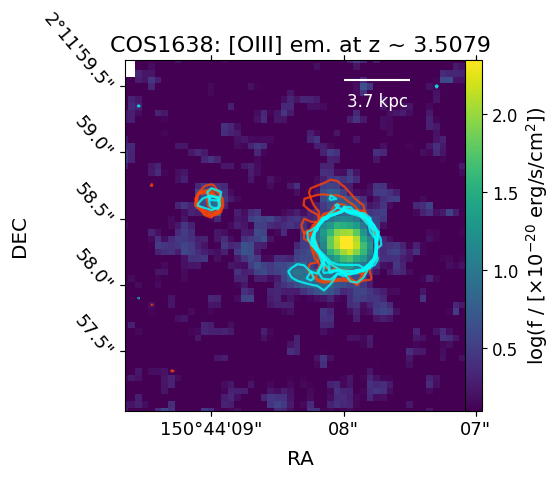}}
\includegraphics[width=0.23\textwidth, trim=0mm 0mm 0mm 4mm]{{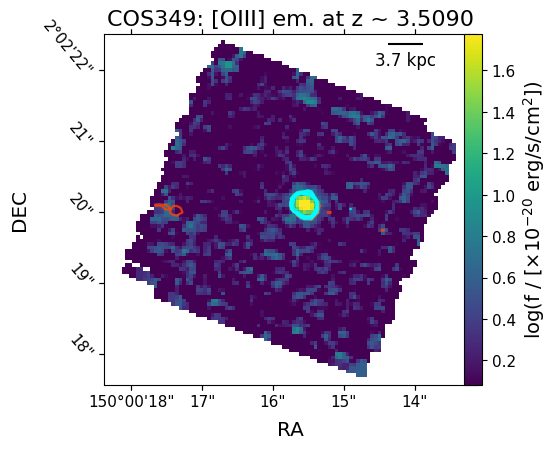}}
\includegraphics[width=0.23\textwidth, trim=0mm 4mm 0mm 4mm]{{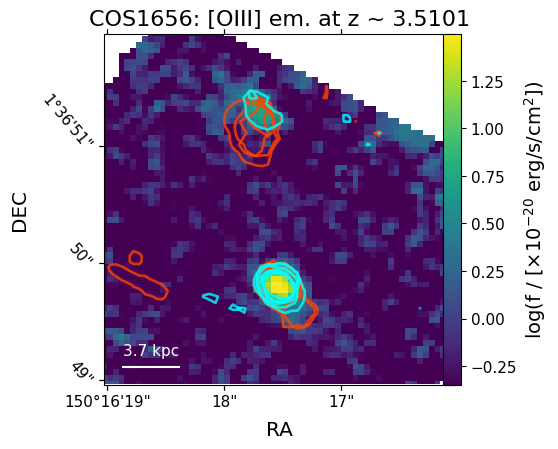}}
\includegraphics[width=0.23\textwidth, trim=0mm 4mm 0mm 4mm]{{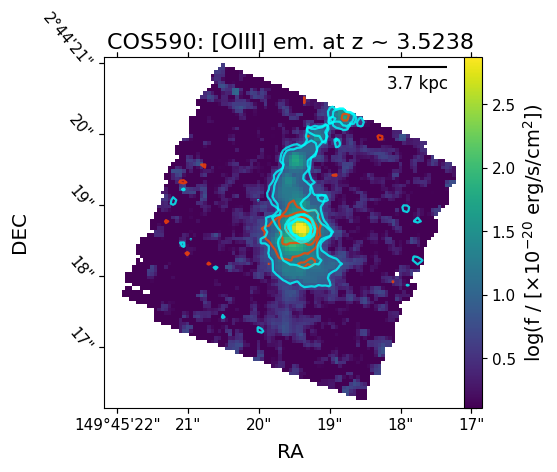}}
\includegraphics[width=0.23\textwidth, trim=0mm 4mm 0mm 4mm]{{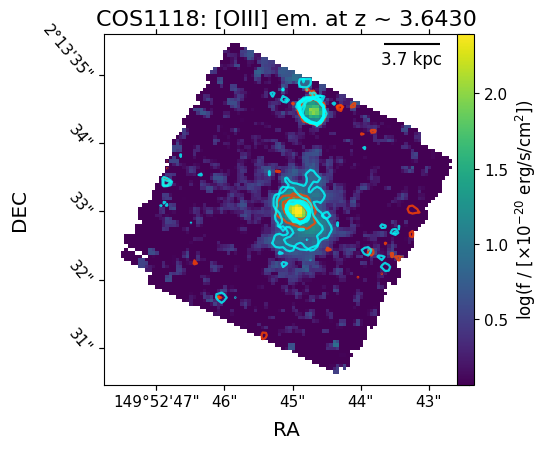}}
\includegraphics[width=0.23\textwidth, trim=0mm 0mm 0mm 4mm]{{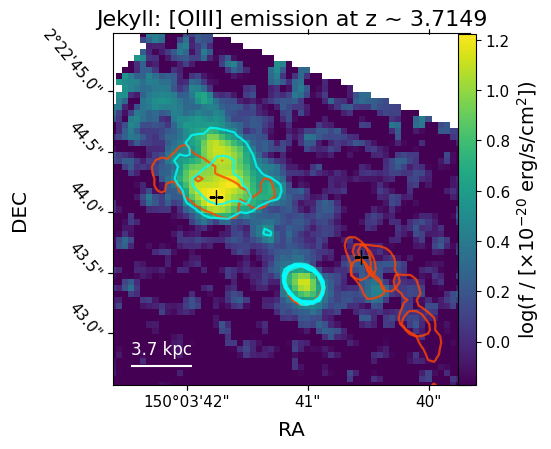}}
\includegraphics[width=0.23\textwidth, trim=0mm 4mm 0mm 4mm]{{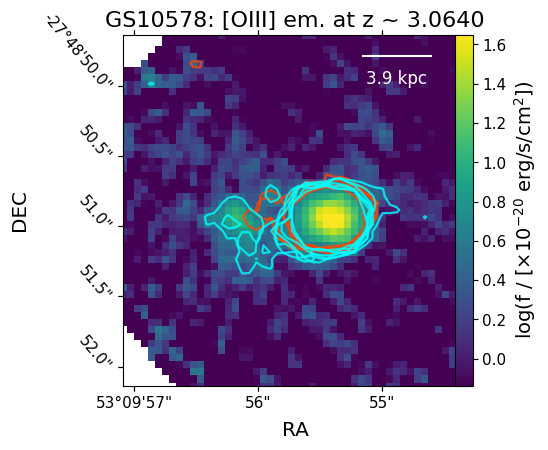}}
\includegraphics[width=0.23\textwidth, trim=0mm 4mm 0mm 4mm]{{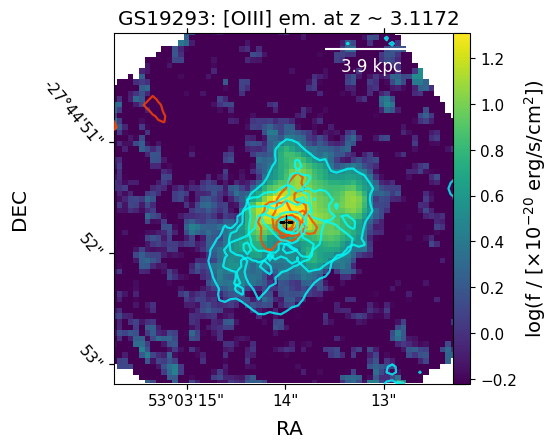}}
\includegraphics[width=0.23\textwidth, trim=0mm 4mm 0mm 4mm]{{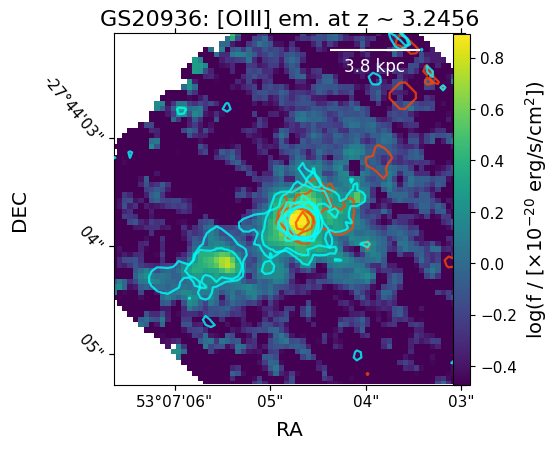}}
\includegraphics[width=0.23\textwidth, trim=0mm 0mm 0mm 4mm]{{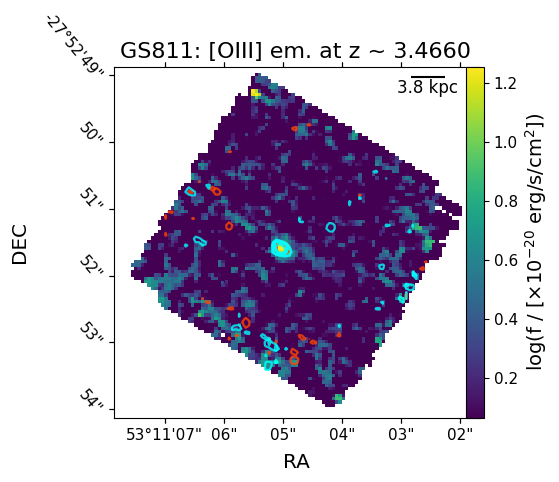}}
\includegraphics[width=0.23\textwidth, trim=0mm 4mm 0mm 4mm]{{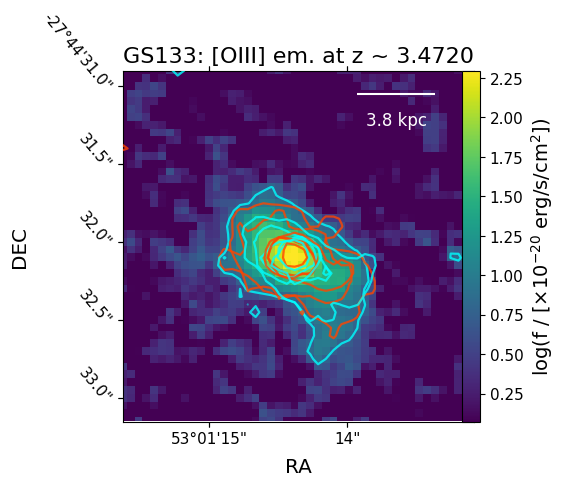}}
\includegraphics[width=0.23\textwidth, trim=0mm 4mm 0mm 4mm]{{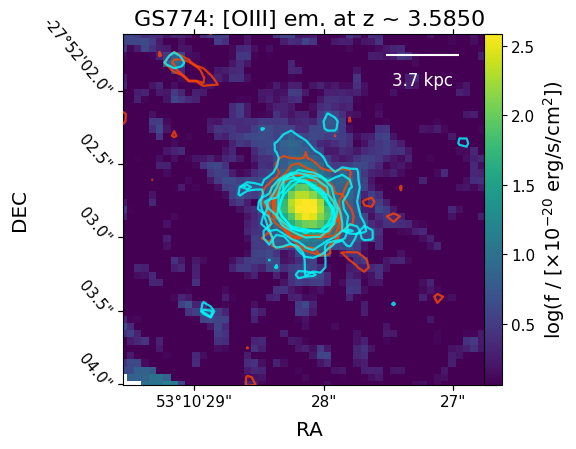}}
\includegraphics[width=0.23\textwidth, trim=0mm 4mm 0mm 4mm]{{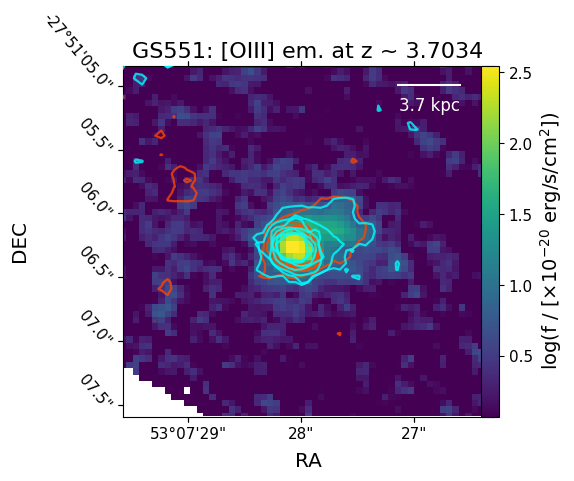}}
\includegraphics[width=0.23\textwidth, trim=0mm 4mm 0mm 4mm]{{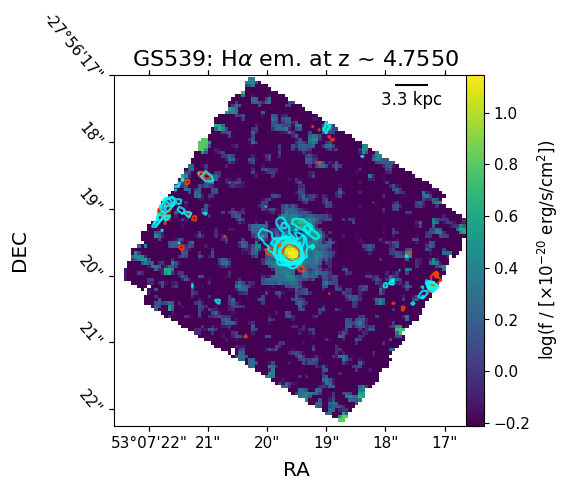}}
\includegraphics[width=0.23\textwidth, trim=0mm 4mm 0mm 4mm]{{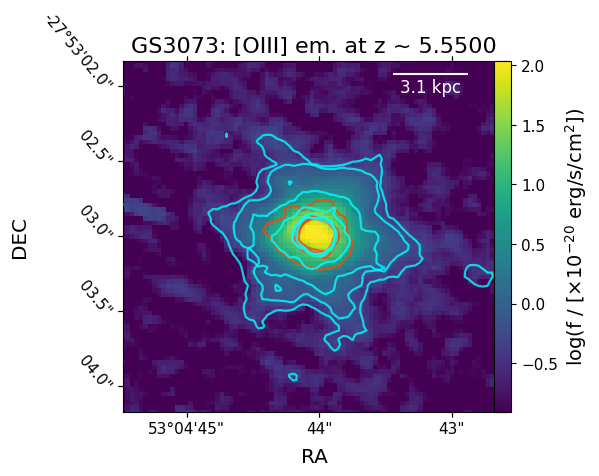}}

\caption{\oiii and \ha maps of the close environments of GA-NIFS AGN systems. They trace the emitting gas observed with NIRSpec IFS in the velocity channels $v\in [-100,100]$~\kms with respect to the AGN systemic redshift. Contours represent regions where the line emission reaches a significance level of 5$\sigma$. Each contour corresponds to emission from a specific velocity bin of 200~\kms, spanning the full velocity range from -1100 to +1100~\kms. Cyan contours indicate blueshifted gas, while orange contours mark redshifted gas.
Cross symbols identify the AGN position in more complex (interacting) systems. All but GS3073 maps were obtained from NIRSpec cubes with 0.05\arcsec/pixel; for GS3073 we opted for the use of the finest 0.03\arcsec/pixel sampling to show the disturbed morphology due to very nearby companions (see also Fig. A1 in \citealt{Ji2024gs3073}). All images are oriented north up, with east to the left. }\label{fig:cutouts}
\end{figure*}

Therefore, the AGN sample studied in this paper consists of 16 systems at $2.0<z<5.6$, with bolometric luminosities in the range L$_{\rm bol}$ $= 10^{45}-10^{47}$ \ergs (Circosta et al., in prep.). The complete list of targets is presented in Table \ref{tab:log}. The \oiii$\lambda$5007 emission line maps for the entire sample are presented in Fig. \ref{fig:cutouts} and  discussed in the next sections.

%\bigskip

\section{Observations and data reduction}\label{sec:observations}

\subsection{NIRSpec IFS data}

The targets presented in this work were observed as part of the NIRSpec IFS GTO survey GA-NIFS (e.g. \citealt{Perna2023IAU}), under programmes \#1216 (PI: Kate Isaak) and \#1217 (PI: Nora Luetzgendorf). Information on the NIRSpec observations is reported in Table \ref{tab:log}. 

The observational design was chosen in order to cover all main optical lines from \hb to  \sii$\lambda \lambda$6716,31 doublet lines with a high spectral resolution (R $\sim 2700$) at $z=3-6$. Therefore, for all but two targets, we used the grating-filter pair G235H/F170LP. The source GS3073, at $z = 5.55$, was observed with G395H/F290LP; then, GS539, at $z= 4.76$, was observed with both G235H/F170LP and G395H/F290LP configurations. Next, COS2949 was observed with G235H/F170LP; we note that it was included in the GA-NIFS survey because it is listed as a $z=3.571$ AGN with a spectroscopic redshift classified as `secure' in the COSMOS catalogues. 
However, NIRSpec observations have revealed that COS2949 is at $z=2.0478$. For this reason, the NIRSpec wavelength range only covers the \ha complex on its bluest side, missing the \oiii and \hb emission lines. 

Exposure times range from $\lesssim 1$ hour for X-ray-selected AGNs, to $\sim 4-5$ hours for the sources originally selected as (fainter) star-forming systems. Therefore, the quality of the data in our sample is not homogeneous. However, all targets have well detected emission lines, and display plenty of faint companions in the NIRSpec field of views (FOVs).

Raw data files were downloaded from the MAST archive and subsequently processed with the JWST Science Calibration pipeline version 1.8.2 under CRDS context jwst\_1068.pmap. We made several modifications to the default three-stages reduction to increase data quality, which are described in detail by \cite{Perna2023} and briefly reported here. We patched the pipeline to avoid over-subtraction of elongated cosmic ray artefacts during the Stage1. The individual count-rate frames were further processed at the end of the pipeline  Stage1, to correct for different zero-level in the individual (dither) frames, and subtract the vertical 1/f noise. In addition, we made the following corrections to the *cal.fits files after Stage 2: we masked pixels at the edge of the slices (two pixels wide) to conservatively exclude pixels with unreliable sflat corrections; we removed regions affected by leakage from failed open shutters; finally, we used a modified version of LACOSMIC (\citealt{vanDokkum2001}) to reject strong outliers before the construction of the final data cube (\citealt{DEugenio2024NatAs}). 
The final cubes were combined using the drizzle method with pixel scales of 0.05\arcsec and 0.03\arcsec, for which we used an official patch to correct a known bug. The smaller pixel scale is required to better characterise the closest dual AGNs, fully exploiting the spatial resolution of the data (\citealt{DEugenio2024NatAs}). 
The astrometric alignment of the NIRSpec data was computed relative to HST images that have been registered to Gaia DR3.

\subsection{MUSE data}

Two of the 16 systems in the GA-NIFS AGN sub-sample have archival VLT/MUSE data, from the MUSE-WIDE (\citealt{Urrutia2019}) and the MUSE-UDF (\citealt{Bacon2017}) surveys. These observations cover the rest-frame UV spectra of GS551 and GS10578, respectively.  

Reduced MUSE datacubes were downloaded from the MUSE-WIDE cut-out service\footnote{\url{https://musewide.aip.de/cutout/}}, for the source GS551, and from the ESO archive, for GS10578. Their observational log is reported in Table \ref{tab:log}. Because the MUSE spectrograph does not operate in vacuum and the wavelength calibration is in standard air\footnote{CTYPE3 = 'AWAV' in the headers of MUSE cubes, corresponding to 'wavelength air'.} (\citealt{Weilbacher2020}), we converted them to vacuum before the fit analysis, consistent with NIRSpec. As for the  NIRSpec observations, the astrometric alignment of the MUSE data was computed relative to HST images registered to Gaia DR3.

\subsection{X-ray data}
X-ray data offer complementary information for assessing AGN activity and constraining the physical properties of these systems. 
The X-ray properties of the sample, especially their X-ray flux measurements, were obtained from the catalogues of the X-ray deep fields COSMOS-Legacy survey \citep{civano2016,Marchesi2016} and the CDFS \citep{liu2017_cdfs} for COSMOS and GOODS-S targets, respectively.  
We further used the full mosaics of COSMOS-Legacy (\citealt{civano2016}) and CDFS \citep{liu2017_cdfs} to estimate the upper limit 0.3--7 keV flux of the systems that are undetected in the X-rays. 
Moreover, we used one of the single exposures of the COSMOS-Legacy survey to study the X-ray morphology of COS1638  (obsID 15253, corresponding to the one in which our target is observed in the best conditions). We downloaded the raw data from the \chandra archive and reduced them with CIAO v4.15 \citep{fruscione2006} through the {\tt chandra\_repro} tool. The morphological analysis of COS1638 is presented in Sect. \ref{sec:xray}.

\subsection{VIMOS data}

In this work, we also analysed the rest-frame UV spectra of 5 X-ray detected, type 2 AGNs at $z\sim 3$, observed as part of the survey ``VANDELS: a VIMOS survey of the CDFS and UDS fields'' (\citealt{Mclure2018}) and previously identified by \citet{Saxena2020} and \citet{Mascia2023}. They were observed using the MR grism and the GG475 filter, with a 1\arcsec\ slit width and a slit length of 10\arcsec\ oriented east-west on the sky. The total exposure time was $\approx 40$ hours per target. This setup provides a wavelength coverage of 4800–10000~$\AA$ with a nominal resolution of R~$= 580$, corresponding to a velocity resolution of approximately 500~\kms. The spectroscopic analysis presented in this work takes advantage of the VANDELS DR4 fully reduced spectra (\citealt{Garilli2021}), downloaded from the ESO archive\footnote{\url{https://archive.eso.org}}.

One of these VANDELS AGN, GS133, was also analysed as part of our GA-NIFS survey, with its UV spectrum thoroughly analysed in \cite{Perna2024gs133}. The five VANDELS targets (discussed in Sect. \ref{sec:uvdiagnostics}) enhance our investigation by contributing to the assessment of the reliability of UV diagnostic diagrams for identifying AGNs at high redshifts.

\section{Data analysis}\label{sec:analysis}

\subsection{Identification of line emitters in NIRSpec data}

To search for potential companion galaxies and line emitters near our AGN systems observed with JWST, we performed a spectral scan around the \oiii$\lambda$5007 and \ha emission lines. Specifically, we explored a velocity range from $\sim -1000$ to +1000~\kms relative to the systemic velocity of the AGN. The \oiii line is generally preferred due to its relative brightness compared to other optical lines, which aids in the clear identification of nearby emitters. However, in heavily obscured AGNs where \oiii is faint (e.g. GS539; \citealt{Parlanti2024aless}), \ha was used as an alternative. For COS2949, \ha was analysed instead of \oiii due to limited wavelength coverage. 

Figure \ref{fig:cutouts} shows emission line cutouts for each target, obtained integrating over the range $v\in [-100,+100]$~\kms, with contours tracing the 5$\sigma$-emission in 200~\kms-wide bins at lower and higher velocities (up to $\pm 1100$~\kms). 
Many of our 16 AGNs are situated in complex environments, showing signs of gravitational interactions (e.g. COS590, COS2949, COS1638), and several \oiii emitters within a few hundred \kms of the central AGN. In particular, close emitters are identified in 9 systems: COS590, COS1638, COS1656, COS2949, GS551, GS3073 (see also \citealt{Ubler2023}), GS20936, Jekyll (\citealt{Perez-Gonzalez2024}), and GS10578 (\citealt{DEugenio2024NatAs}). A bright M star is identified in the vicinity of COS1118 (in the upper part of the NIRSpec FOV, at RA = 09:59:31, DEC = 02:13:34.3).

\subsection{NIRSpec spectroscopic analysis}

To investigate the nature of the line emitters in the vicinity of our GA-NIFS AGN, we extracted integrated spectra from circular apertures with $r=0.1\arcsec\ $ centred at the peak position of all identified emitters (comprising the bright AGN), hence from regions broadly corresponding to the spatial resolution element of NIRSpec IFS observations (\citealt{DEugenio2024NatAs}). 

We modelled their integrated spectra with a combination of Gaussian profiles. 
We fitted the most prominent gas emission lines by using the Levenberg-Marquardt least-squares fitting code CAP-MPFIT
(\citealt{Cappellari2017}). In particular, we modelled the \ha and \hb  lines, the \oiii$\lambda\lambda$4959,5007, \nii$\lambda\lambda$6548,83, and \sii$\lambda\lambda$6716,31
doublets applying a simultaneous fitting procedure (\citealt{Perna2023}), so that all line features corresponding to a given kinematic component have the same velocity centroid and full width at half maximum (FWHM). During the fit, we removed all Gaussian components, whose amplitude is characterised by a signal-to-noise ratio of S/N~$< 2$. The modelling of the \ha and \hb BLR emission in type 1 AGN sources required the use of broken power-law components (e.g. \citealt{Cresci2015a}): they are preferred to a combination of extremely broad Gaussian profiles, because the former tend to minimise the degeneracy between narrow line region (NLR) and BLR emission. Finally, we used the theoretical model templates of \cite{Kovacevic2010} to reproduce the \feii emission in the wavelength region $4000-5500$~\AA\ (see \citealt{Perna2017a} for details). The final number of kinematic components used to model the spectra is derived on the basis of the Bayesian information criterion (BIC, \citealt{Schwarz1978}). The continuum emission, when detected, is modelled with a low-order polynomial function, and subtracted prior to fitting the emission lines. For GS10578, which shows strong stellar continuum and absorption line features, we used the model obtained by \citet{DEugenio2024NatAs}.

Following \cite{Ubler2023}, the noise level was obtained from the ERR extension of the NIRSpec data cubes, after re-scaling it with a measurement of the standard deviation in the integrated spectrum in regions free of line emission to take into account correlations due to the non-negligible size of the PSF relative to the spaxel. 
The uncertainties in the emission line properties reported in this work were measured using Monte Carlo simulations (following e.g. \citealt{Perna2015a}). We collected 50 mock spectra using the best-fit final models, added random noise, and
fit them. The errors were then calculated by taking the range that contains 68.3\% of values evaluated from the obtained distributions
for each measurement.

\begin{figure}[!h]
\centering
\includegraphics[width=0.23\textwidth, trim=0mm 7mm 0mm 7mm]{{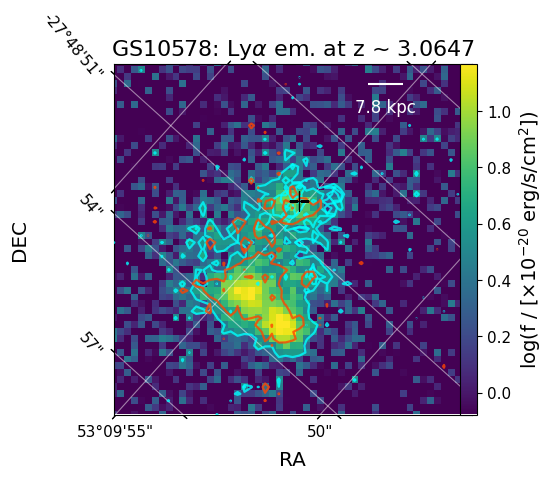}}
\includegraphics[width=0.23\textwidth, trim=0mm 7mm 0mm 7mm]{{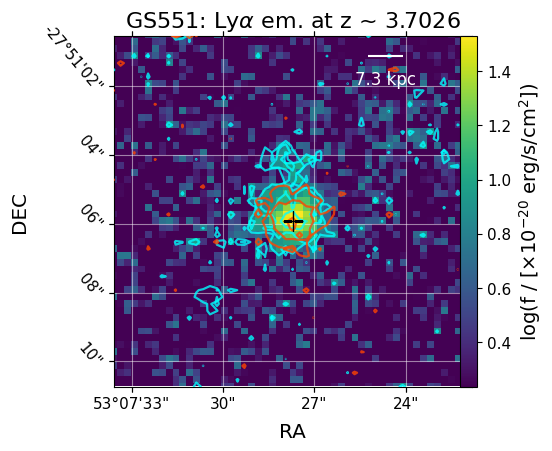}}

\caption{ \lya maps of the close environments of GS10578 (left) and GS551 (right). They trace the emitting gas with $v\in [-200,200]$~\kms observed with MUSE. The cyan (orange) contours indicate the \lya emission at 4$\sigma$ in velocity bins of 400~\kms, from -1400 to + 1400~\kms. The cross symbols identify the (NIRSpec) AGN positions. The coordinate grids highlight the different rotation of MUSE data with respect to the standard north-east up-left configuration. }\label{fig:MUSEcutouts}
\end{figure}

\begin{figure*}[!h]
\centering
\includegraphics[width=0.83\textwidth, trim=0mm 8mm 0mm 8mm]{{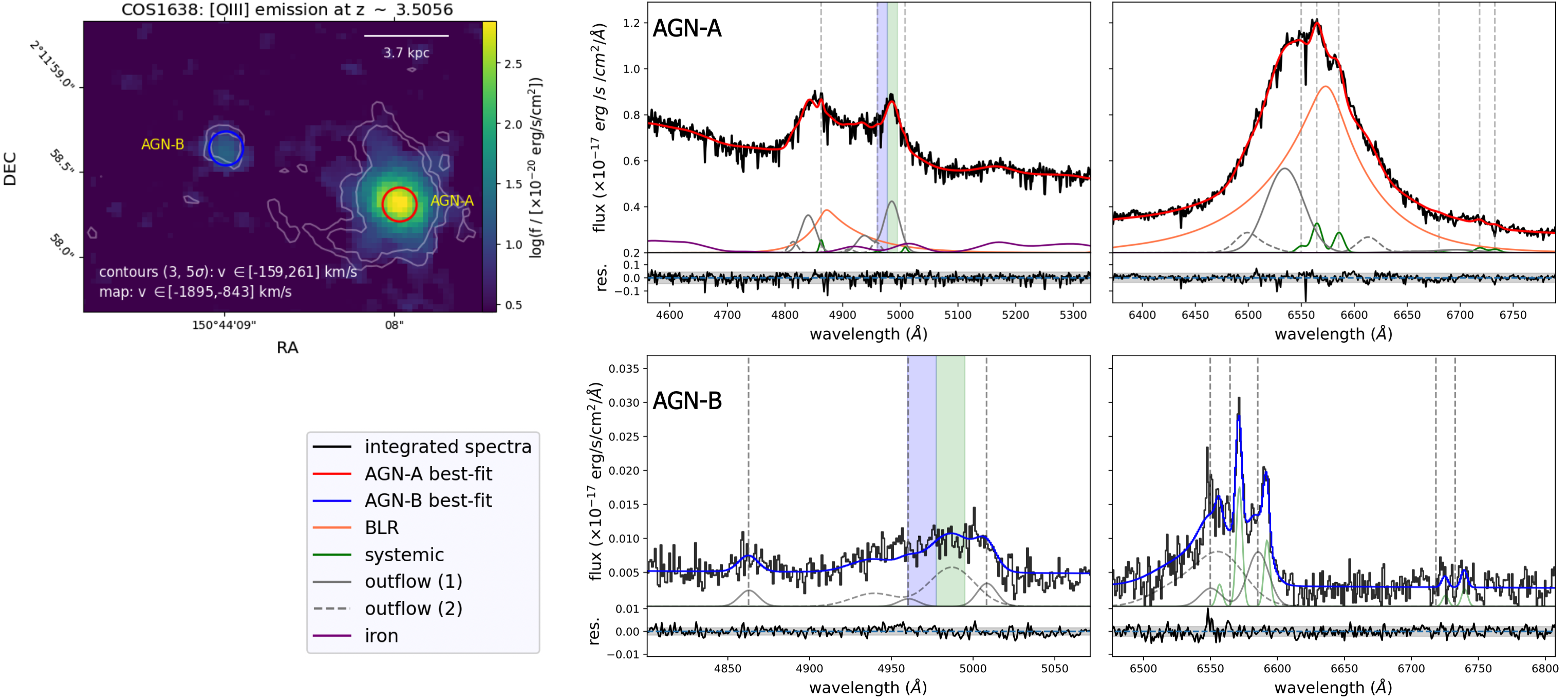}}

\caption{ \oiii map (top-left) and integrated spectra of the dual AGN system associated with COS1638. The \oiii map is obtained by integrating the \oiii emission over specific velocity ranges (as labelled), in order to highlight the primary (AGN-A) and secondary (AGN-B) nuclei, together with additional structures at the same redshift of the active galaxies. The panels on the right show the integrated spectra of the primary (top) and secondary (bottom) AGNs, extracted from the circular regions labelled in the map. The legend displays the individual Gaussian components and best-fit curves shown in the spectra; best-fit total profiles are represented in red for AGN-A and in blue for AGN-B. Residuals, defined as the difference between the observed data and the best-fit model, are displayed at the bottom of each panel; average $\pm 3\sigma$ uncertainties on the data are also reported with horizontal grey shaded regions. The vertical blue and green shaded areas correspond to the velocity channels used to generate the \oiii map and contours in the top-left panel, respectively. Both targets present broad lines due to powerful outflows. }\label{fig:cos1638spec}
\end{figure*}

\subsection{Identification of line emitters in MUSE data}

We used MUSE data cubes to identify possible companions and line emitters in the vicinity of GS10578 and GS551. 
This analysis followed the methodology previously applied to NIRSpec data but extended the search to a larger area beyond the relatively small NIRSpec FOV ($\sim 3\arcsec\times3\arcsec$).
Specifically, we conducted a spectral scan around the \lya emission line, the brightest line in the UV regime; we covered a velocity range from -1400 to +1400~\kms relative to the AGN systemic velocity, with 400~\kms bins. The broader velocity range and larger binning reflect the lower S/N in the MUSE data cubes.

Figure \ref{fig:MUSEcutouts} shows the emission line cutouts for both targets, obtained integrating over the range $v\in [-200,200]$~\kms, with contours tracing the 4$\sigma$-emission in bins at $v<-200$~\kms and $v>200$~\kms. They show extended nebulae surrounding the central X-ray detected AGNs. Moveover, GS10578 shows two close \lya emitters, brighter than the one of the central AGN. GS551 shows a faint \lya emitter at 20~kpc towards the south-east, and diffuse emission towards the north and north-west.

\subsection{Spectral fit of MUSE data}

The MUSE spectroscopic data were analysed following the same general procedures described in the previous section. In particular, we extracted the spectra of individual emitters by integrating regions optimised to increase the S/N for each source.  
The fit analysis required the use of a single Gaussian profile for the lines \heii $\lambda$1640, \civ$\lambda\lambda$1549, 1550, and \ciii$\lambda$1907, 1909 (with the latter two modelled taking into account their doublet nature), which are the lines required for the classification of AGNs and SFGs according to the UV diagnostics proposed by \cite{Feltre2016} and \cite{Nakajima2022}.
Also for these targets, the outcomes of the spectroscopic analysis are discussed in the following sections.

\begin{figure*}[!h]
\centering
\includegraphics[width=0.99\textwidth, trim=0mm 8mm 0mm 8mm]{{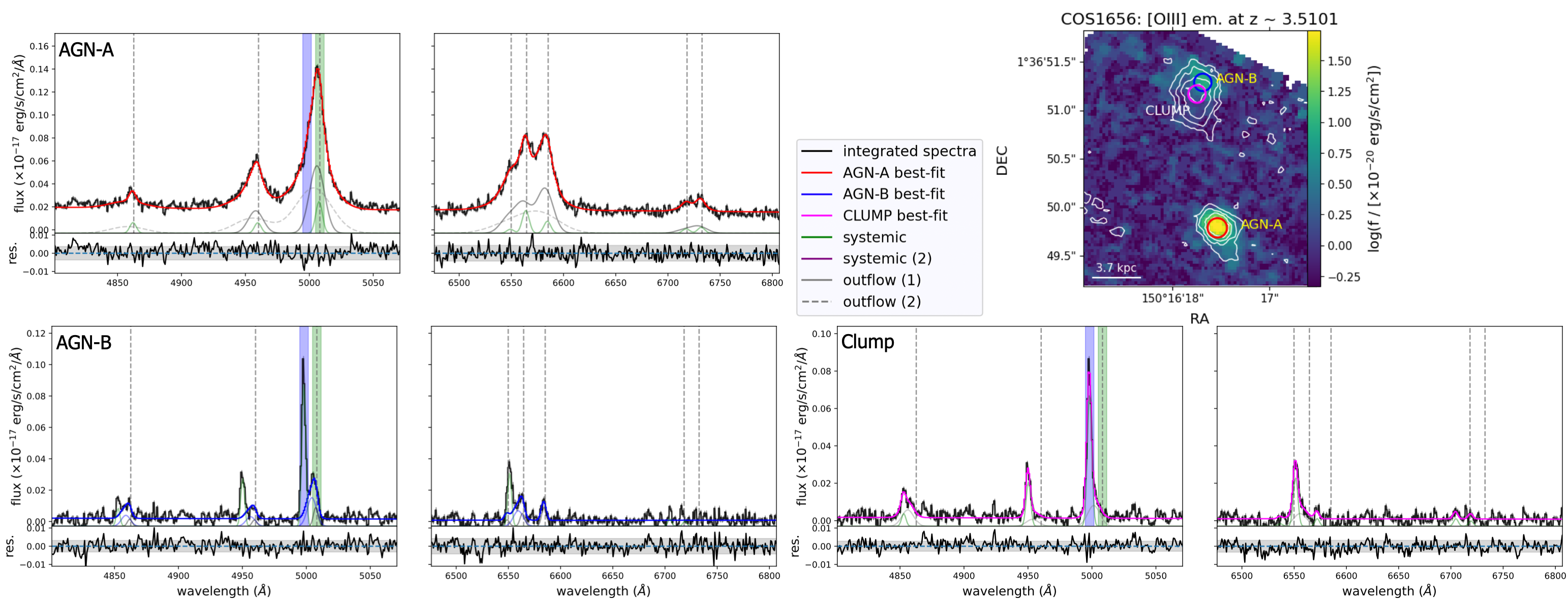}}

\caption{ \oiii map (top-right) and integrated spectra of the dual AGN system and a faint clump associated with COS1656. The \oiii map is obtained by integrating the \oiii emission over a velocity range from $-200$ to $+200$~\kms, in order to highlight the primary (AGN-A) and secondary (AGN-B) nuclei; the contours trace the emission at $v \in [-800,-600]$ associated with a clump close to AGN-B. The remaining panels show the integrated spectra of the primary (top-left) and secondary (bottom-left) AGNs, as well as of the clump (bottom-right), extracted from the circular regions labelled in the map. The legend display the individual Gaussian components and best-fit curves shown in the spectra; best-fit total profiles are represented in red for AGN-A, in blue for AGN-B, and in magenta for the clump. Residuals, defined as the difference between the observed data and the best-fit model, are displayed at the bottom of each panel; average $\pm 3\sigma$ uncertainties on the data are also reported with grey shaded regions. The vertical green and blue shaded areas at $\sim 5007\AA$ mark the channels used to generate the \oiii map and contours in the top-right panel, respectively. }\label{fig:cos1656spec}
\end{figure*}

\begin{figure*}[!h]
\centering
\includegraphics[width=0.99\textwidth, trim=0mm 8mm 0mm 0mm]{{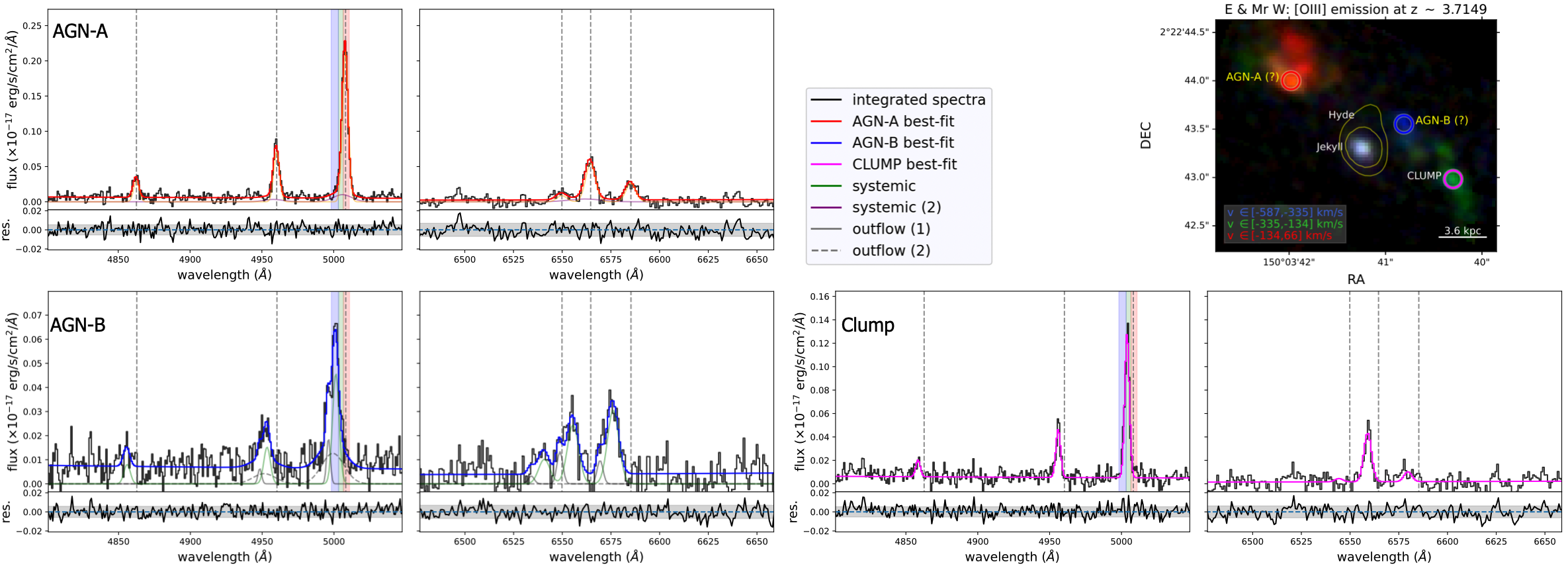}}

\caption{ \oiii map (top-right) and integrated spectra of the dual AGN system and a faint clump associated with Jekyll \& Hyde. The \oiii map is obtained by integrating the line emission over specific velocity ranges (as labelled), in order to highlight the primary (Eastfield, AGN-A) and secondary
(Mr West, AGN-B) nuclei, together with additional structures at the same redshift of the active galaxies. The remaining panels show the integrated spectra of the primary (top-left) and secondary (bottom-left) AGNs, as well as of a compact clump between the two brightest sources (bottom-right). The legend display the individual Gaussian components and best-fit curves shown in the spectra. Residuals, defined as the difference between the observed data and the best-fit model, are displayed at the bottom of each panel, with grey shaded regions indicating the average $\pm 3\sigma$ uncertainties on the data. The vertical shaded areas at $\sim 5007\AA$ mark the channels used to generate the \oiii three-colour map. }\label{fig:jekyllspec}
\end{figure*}

\begin{figure*}[!h]
\centering
\includegraphics[width=0.99\textwidth, trim=0mm 8mm 0mm 8mm]{{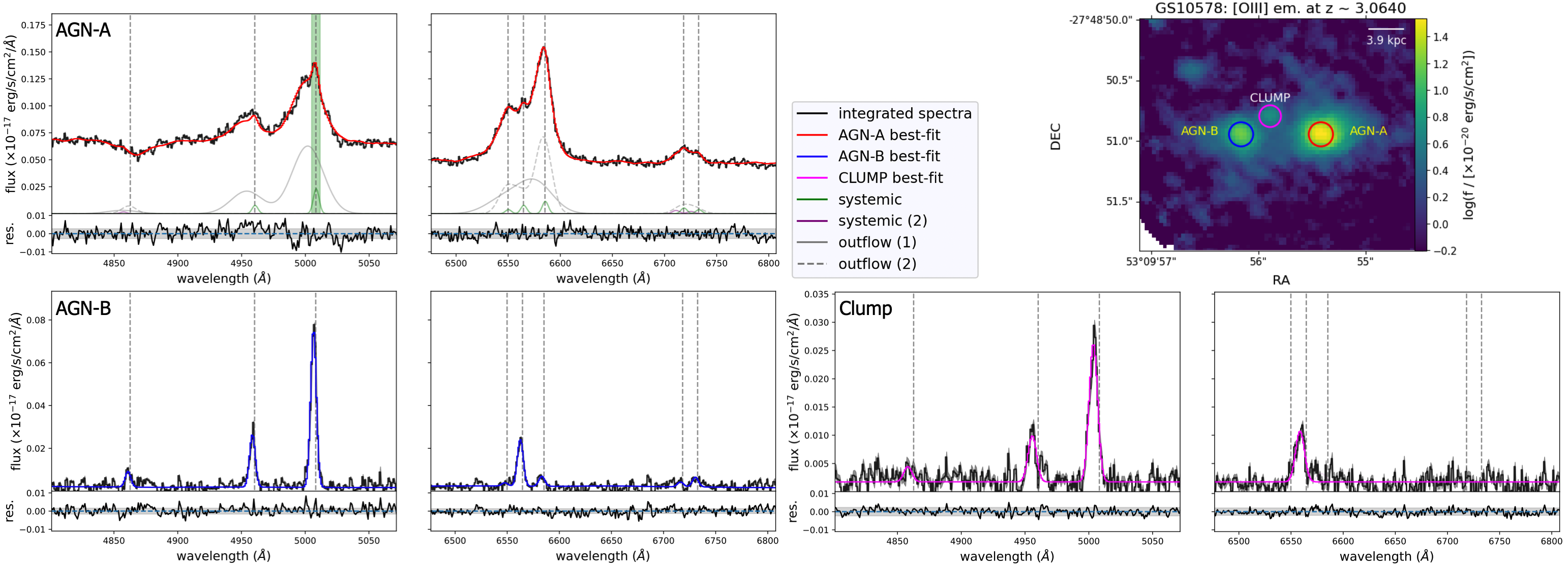}}

\caption{ \oiii map (top-right) and integrated spectra of the dual AGN system and a faint clump associated with GS10578. The \oiii map is obtained by integrating the \oiii emission over a velocity range from $-200$ to $+200$~\kms, in order to highlight the primary (AGN-A) and secondary (AGN-B) nuclei, together with additional structures at the same redshift of the active galaxies. The remaining panels show the integrated spectra of the primary (top-left) and secondary (bottom-left) AGNs, as well as of a compact clump between the two brightest sources (bottom-right). The legend display the individual Gaussian components and best-fit curves shown in the spectra. Residuals, defined as the difference between the observed data and the best-fit model, are displayed at the bottom of each panel, with grey shaded regions indicating the average $\pm 3\sigma$ uncertainties on the data. The vertical green shaded area at $\sim 5007\AA$ mark the channels used to generate the \oiii map. }\label{fig:gs10578spec}
\end{figure*}

\begin{figure*}[!h]
\centering
\includegraphics[width=0.99\textwidth, trim=0mm 8mm 0mm 0mm]{{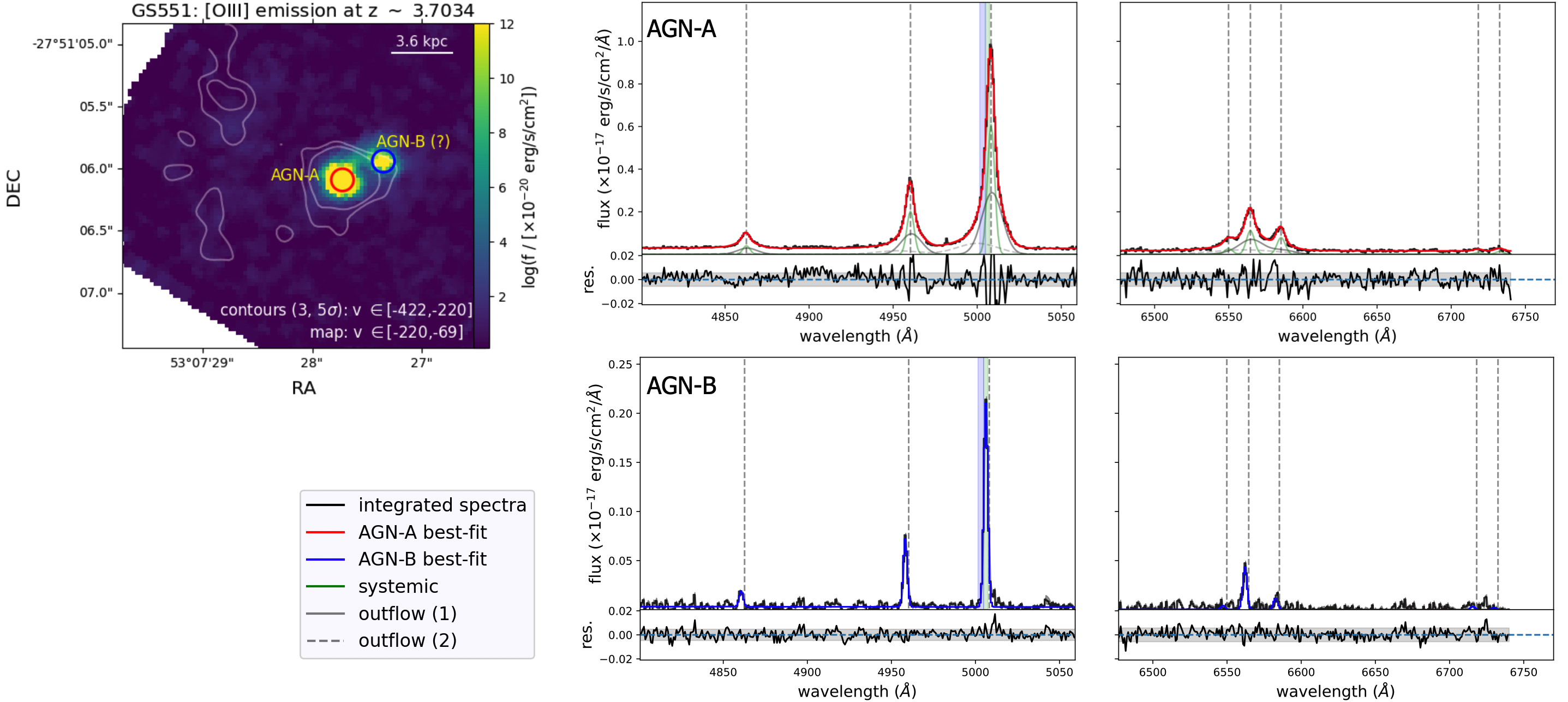}}

\caption{ \oiii map (top-right) and integrated spectra of the candidate dual AGN system associated with GS551. The \oiii map is obtained by integrating the \oiii emission over a specific velocity range (as labelled), in order to highlight the primary (AGN-A) and secondary (AGN-B) nuclei, together with additional structures at the same redshift of the active galaxies. The remaining panels show the integrated spectra of the primary (top-right) and secondary (bottom-right) AGNs. The legend display the individual Gaussian components and best-fit curves shown in the spectra; best-fit total profiles are represented in red for AGN-A, and in blue for AGN-B. Residuals, defined as the difference between the observed data and the best-fit model, are displayed at the bottom of each panel, with grey shaded regions indicating the average $\pm 3\sigma$ uncertainties on the data. The vertical green shaded area at $\sim 5007\AA$ marks the channels used to generate the \oiii map, while the blue shaded area marks those associated with extended emission marked with contours overlapping the map.}\label{fig:gs551spec}
\end{figure*}

\section{Results}

Figure \ref{fig:cutouts} illustrates that nine out of the 16  GA-NIFS targets at $z=2-6$ have close \oiii and \ha line emitters within a $\sim 20 \times 20$ kpc$^2$ area surrounding the main source.  
Similarly, Fig. \ref{fig:MUSEcutouts} demonstrates that both targets with available MUSE observations contain \lya emitters at distances extending to several tens of kiloparsecs. 
In this section we examine five of these systems, which are those with newly identified companions that likely host accreting SMBHs. Their main properties are summarised in Table~\ref{tab:DAGN}, and discussed in detail in Appendix \ref{sec:Atargets}. 
For clarity, we refer to the brighter active galaxy in each system as the `primary' AGN (AGN-A) and the fainter one as the `secondary' AGN (AGN-B). Additionally, in one system, we identified a further line emitter classified as the `tertiary' AGN, labelled AGN-C. 
These five systems also present fainter emitters in addition to those classified as AGNs. For simplicity, we refer to these sources as 'clumps'; they may correspond to small satellite galaxies (e.g. near COS1656 and GS10578) or to star-forming regions in tidal tails or debris resulting from gravitational interactions between galaxies (e.g. near Jekyll), associated with faint continuum emission.

To assess the nature of each line emitter and determine whether their emission originates from AGN activity, star formation, or shock excitation, we employed a series of emission-line diagnostics. These diagnostics also allowed us to evaluate whether the observed line emission is due to external ionisation from the primary AGN or if it indicates an intrinsic ionising source, such as a secondary AGN within the companion galaxy. For COS1638, we also present tentative evidence of X-ray emission from the secondary AGN, analysing archival \chandra data. Finally, rest-frame optical and UV observations were complemented with available multi-wavelength information to constrain the stellar masses of the primary and secondary AGNs. The specific diagnostics used to characterise our targets are described in the following subsections. A detailed spatially resolved study of all GA-NIFS AGNs will be presented in forthcoming works (Bertola et al., in prep; Venturi et al., in prep.).

\subsection{Optical emission lines and BPT diagnostics}\label{sec:resultsOPT}

Figures \ref{fig:cos1638spec},  \ref{fig:cos1656spec}, \ref{fig:jekyllspec}, \ref{fig:gs10578spec}, and \ref{fig:gs551spec} show the NIRSpec spectra of the line emitters along with their best-fit models. For each system, we also provide a zoomed-in \oiii map to illustrate the spatial locations of these emitters. All maps were generated from data cubes with a 0.03\arcsec/pixel resolution, to fully leverage the angular resolution of NIRSpec. Optical spectra were extracted from cubes with a 0.05\arcsec/pixel scale for most targets, optimising the S/N. However, for the clump near GS10578, the clump and AGN-B close to COS1656 that overlap on the LOS, and the X-ray AGN GS551 and its close companion, we extracted the spectra from cubes with the finest sampling (0.03\arcsec/pixel) to minimise PSF contaminations.

In all figures, the red colour is used to identify the primary AGN both in spectra (red best-fit curves) and maps (red circles). The secondary, fainter AGN is shown in blue. Any additional sources in the vicinity are marked in magenta for clarity. A detailed description of each system is reported in Appendix \ref{sec:Atargets}. Here we report the main outcomes of the spectroscopic analysis in terms of emission-line ratios and the presence of strong outflows.

For at least one target, AGN-B in COS1638, the detection of a prominent \oiii outflow with velocities reaching several thousand \kms (Table~\ref{tab:fit}) serves as compelling evidence of AGN activity. Such extreme velocities are inconsistent with starburst-driven outflows (e.g. \citealt{Cicone2016, Carniani2024}) and are instead indicative of energetic AGN winds (e.g. \citealt{Perrotta2019, Villar2020, Tozzi2024}). 
In particular, the \oiii features in COS1638-B are characteristic of AGN classified as `blue outliers' (\citealt{Zamanov2002}), where the peak of \oiii\ is blueshifted relative to the systemic velocity corresponding to the peak of the \ha narrow component. Similar profiles are observed in the secondary AGN of the BR1202--0725 system, reported in \citet{Zamora2024}.

Asymmetric emission line profiles are also observed in other AGN-B, albeit with less pronounced wings that could correspond to weaker outflows (Figs.~\ref{fig:cos1656spec}, \ref{fig:jekyllspec}, \ref{fig:gs551spec}). This emission might result from starburst or AGN winds but could also be due to gravitational interactions between the dual AGN hosts. Additionally, PSF smearing may contribute to the broadening of line profiles, as seen in AGN-B of GS551, and in AGN-B and the clump associated with COS1656. In fact, all newly discovered line emitters lie within 10~kpc from the central primary AGN, and have velocity shifts of a few hundreds of~\kms, indicating that gravitational interactions are probable. Filamentary structures likely associated with tidal tails are also detected in the close surroundings of COS1638 (Fig. \ref{fig:cos1638spec}), Jekyll (Fig. \ref{fig:jekyllspec}), and GS551 (Fig. \ref{fig:gs551spec}). Consequently, except for COS1638 AGN-B, the kinematic signatures alone cannot be reliably used to confirm the presence of multiple AGNs within these systems. To asses the nature of the newly discovered line emitters, we used line ratio diagnostics.

\begin{figure}[!htb]
\centering
\includegraphics[scale=0.58, trim=0mm 7mm 0mm 7mm]{{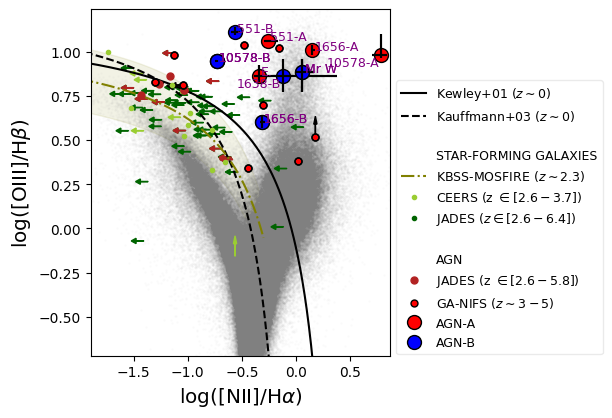}}

\caption{ BPT diagram. Large red and blue circles refer to primary (AGN-A) and secondary (AGN-B) nuclei, respectively. 
All primary and secondary AGNs are above the solid (\citealt{Kewley2001}) and dashed (\citealt{Kauffmann2003}) curves, used to separate purely SFGs (below the curves) from galaxies containing AGNs (above the curves, i.e. the Seyfert region of the BPT). 
The olive curve indicates the locus of $z\sim 2.3$ SFGs from \cite[with intrinsic scatter relative to the median curve in light-olive][]{Strom2017}; additional measurements for $z > 2.6$ SFGs and AGNs are reported with small symbols (see legend and Sect. \ref{sec:opticaldiagnostics} for details). All reported measurements show that the Seyfert area of the BPT is free from contamination by SFGs at any considered redshift, and, therefore, ensure the AGN classification of the newly discovered sources COS1638-B, GS551-B, Eastfield, Mr. West, COS1656-B, and GS10578-B. All AGN-A and B flux ratios are reported in Table \ref{tab:fit}.
}\label{fig:BPT}
\end{figure}

The ionisation source of each AGN is determined using the standard BPT diagnostic diagram (\citealt{Baldwin1981}), shown in Fig. \ref{fig:BPT}. In particular, we report all the line ratios with clear AGN ionisation (AGN-A in red, AGN-B in blue); hence, it does not show those of individual clumps whose line ratios could be associated both with star formation and AGNs, because they have upper limits in the \nii/\ha. We also excluded from the BPT the primary AGN in COS1638, as it is affected by fit degeneracy between BLR and NLR components. For completeness, the individual line ratios of all line emitters in the five systems studied in this work are reported in Table \ref{tab:nirspeclineratios}. 

For reference, the BPT diagnostic in Fig. \ref{fig:BPT} also displays the optical line ratio measurements from the remaining type 2 AGN in GA-NIFS (small red circles), for which spectra will be presented in Bertola et al. (in prep.) and Venturi et al. (in prep.). The BPT also presents other measurements from the literature. In particular, there are low-$z$ SDSS galaxies (small grey points), the median locus of SFGs at $z\sim 2.3$ from \citet[light-olive curve with intrinsic scatter]{Strom2017}, and SFGs and AGNs at $z > 2.6$ recently observed by JWST (from \citealt{Scholtz2023} and \citealt{Calabro2023}).  The  demarcation lines used to separate galaxies and AGNs at $z\sim 0$ from \citet{Kewley2001} and \citet{kauffmann_2003} are also marked in the diagram.

The flux ratios for our primary and secondary AGNs fall well above those of distant ($z \gtrsim 3$) star-forming galaxies observed so far (e.g. \citealt{Sun2023, Sanders2023, Nakajima2023}) and occupy the same region as local Seyfert galaxies and AGNs at cosmic noon, providing strong support for their classification as active nuclei. 
The high \nii/\ha line ratios suggest that these AGNs reside in metal-enriched environments, as expected for massive systems (\citealt{Karouzos2014, Curti2023}); in fact, all our primary AGNs have stellar masses log(M$_*$/M$_\odot$)$~> 9.8$ (see Sect. \ref{sec:stellarmasses}). A more detailed discussion about the use of BPT to identify AGNs at $z\gtrsim 2.6$ is presented in Sect. \ref{sec:opticaldiagnostics}.

%\bigskip

\subsection{UV emission lines and line ratio diagnostics}\label{sec:resultsUV}

\begin{figure*}[!ht]
\centering
\includegraphics[scale=0.57, trim=0mm 0mm 0mm 15mm]{{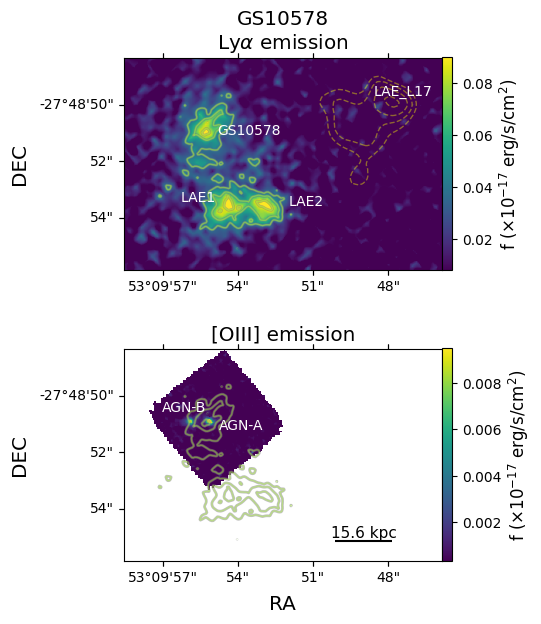}}\hspace{1cm}
\includegraphics[scale=0.5, trim=0mm 0mm 0mm 5mm]{{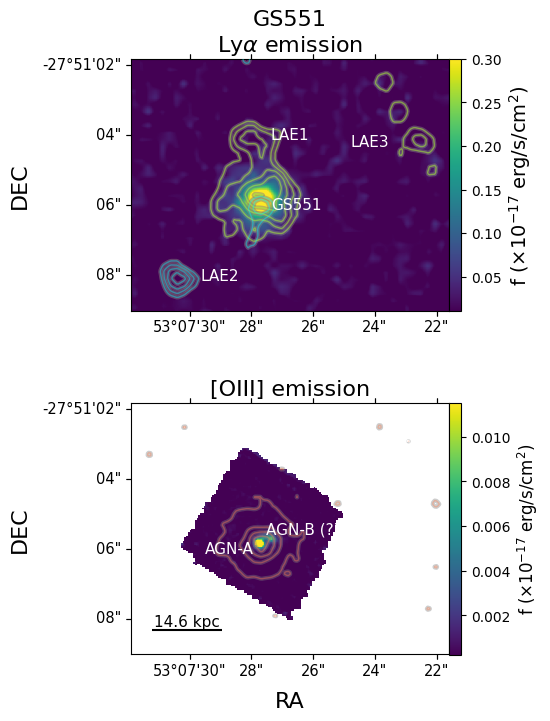}}

\includegraphics[scale=0.57, trim=0mm 7mm 0mm 4mm]{{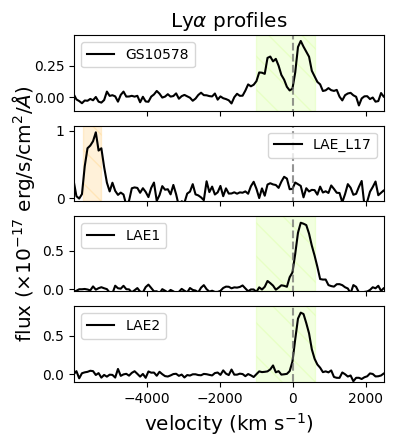}}\hspace{3cm}
\includegraphics[scale=0.57, trim=0mm 7mm 0mm 3mm]{{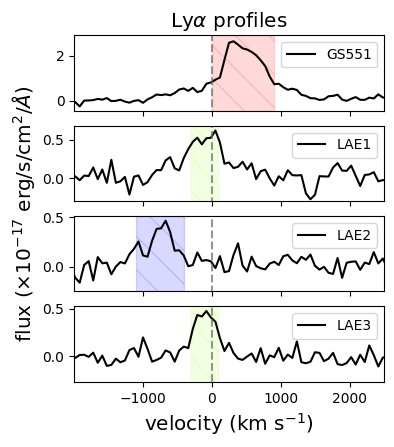}}

\caption{ Comparison between \lya  (top) and \oiii emission maps (middle panels) for GS10578 (left) and GS551 (right), and spectra extracted from invidual \lya emitters (bottom). Both MUSE and NIRSpec images have been sampled with 0.05\arcsec/spaxel, to ease the comparison between UV and optical emission line spatial distributions. The \lya maps are integrated over the emission profile at the location of the X-ray AGN, with overlaid contours indicating emissions at different wavelengths and highlighting nearby \lya emitters. The bottom panels show spectra near the \lya line for each system, in velocity space with respect to the systemic redshifts of the primary AGN; these spectra were extracted from regions with S/N$~>5$. In GS10578, two bright \lya emitters (LAE1 and LAE2) are located $\sim20$~kpc south of the AGN; a \lya halo at slightly lower redshift ($z = 2.99$) previously identified by \citet{Leclercq2017} is also marked with orange contours.
GS551 features three \lya emitters at varying distances up to $\sim30$~kpc; an even more distant \lya emitter located $\sim 80$~kpc to the north-east was also identified by \citet[][see their Fig. 2]{denBrok2020}. 
The contours in the map panels are in steps of $1\sigma$,  for LAE\_L17 and the GS551 companions, and in steps of $3\sigma$ ($10\sigma$) for GS10578 (GS551); all contours start from $5\sigma$ and are obtained integrating over the velocity channels shown in the bottom spectra according to their colours.
These MUSE observations illustrate that GS551 and GS10578 are embedded in dense, complex environments.
}\label{fig:lyaoiiimaps}
\end{figure*}

Figure \ref{fig:lyaoiiimaps} shows a comparison between the \lya emission maps derived from MUSE cubes (top panels) and the \oiii emission observed with NIRSpec (middle panels) for the two GA-NIFS AGNs covered by MUSE observations: GS10578 and GS551. The \lya maps were created by integrating the emission across the line profiles detected at the position of the X-ray AGN (GS10578) and the close companions (GS551). The \lya spectral profiles are reported in the bottom panels of the figure.  
The overlaid contours in the maps mark the emission at different wavelengths, highlighting the presence of the \lya emitters (LAEs) around the central AGN, which spectral lines are reported with different colours in the bottom panels.

While the MUSE data do not resolve individual \oiii emitters identified by NIRSpec, they reveal extended \lya nebulae that encompass both the primary and secondary AGNs, stretching several kiloparsecs from the central  SMBHs.
In GS10578, two prominent \lya emitters are detected approximately 20~kpc south of the central X-ray AGN, and appear nearly as luminous as the primary AGN itself. These two sources, labelled LAE1 and LAE2, have systemic velocities within a few hundreds of~\kms of the GS10578-A systemic redshift. Additionally, these two emitters lie $\sim 50$ kpc (and $\sim 6000$ \kms) from a previously discovered \lya halo identified by \citet[LAE\_L17 in the figure]{Leclercq2017}. 

The MUSE data reveal three additional \lya emitters in the surroundings of GS551: a faint source located $\sim10$~kpc to the north (LAE1), a compact source located $\sim 20$~kpc to the south-east (LAE2), and a third $\sim 30$~kpc to the north-west with a clumpy morphology extending for nearly 20~kpc (LAE3). All of these systems share systemic redshifts within a few hundreds of \kms of the systemic velocity of GS551-A. 

The integrated spectra of the emitters associated with GS10578 and GS551, obtained by summing the spaxels within the outer contours corresponding to S/N = 5, are shown in the bottom panels of Fig.~\ref{fig:lyaoiiimaps}. 
This first qualitative analysis of the archival MUSE data indicate that both GS10578 and GS551 reside in highly complex and dense environments.

%\bigskip
%\bigskip

\begin{figure*}[!htb]
\centering

\includegraphics[scale=0.52, trim=0mm 5mm 0mm 5mm]{{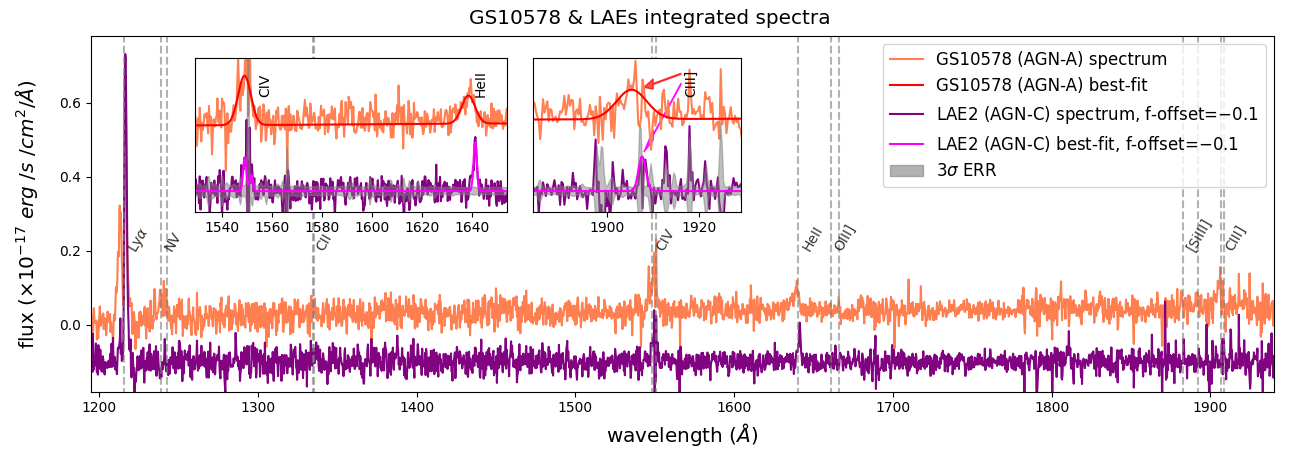}}

\caption{ MUSE spectra of GS10578 (coral) and LAE2 (purple). 
These spectra were obtained by integrating over the regions at S/N$~>4$ in Fig. \ref{fig:lyaoiiimaps}; for visual purposes, the LAE2 spectrum was shifted vertically, as labelled. The zoom-in insets show the same spectra in the vicinity of the lines used in the UV diagnostics used to classify LAE2 as an AGN (Fig. \ref{fig:figUVBPT}), together with 3$\sigma$ errors for the LAE2 spectrum (grey area, mostly highlighting the position of strong sky line residuals).}\label{fig:GS10578musespec}
\end{figure*}

\begin{figure*}[!htb]
\centering

\includegraphics[scale=0.52, trim=0mm 5mm 0mm 5mm]{{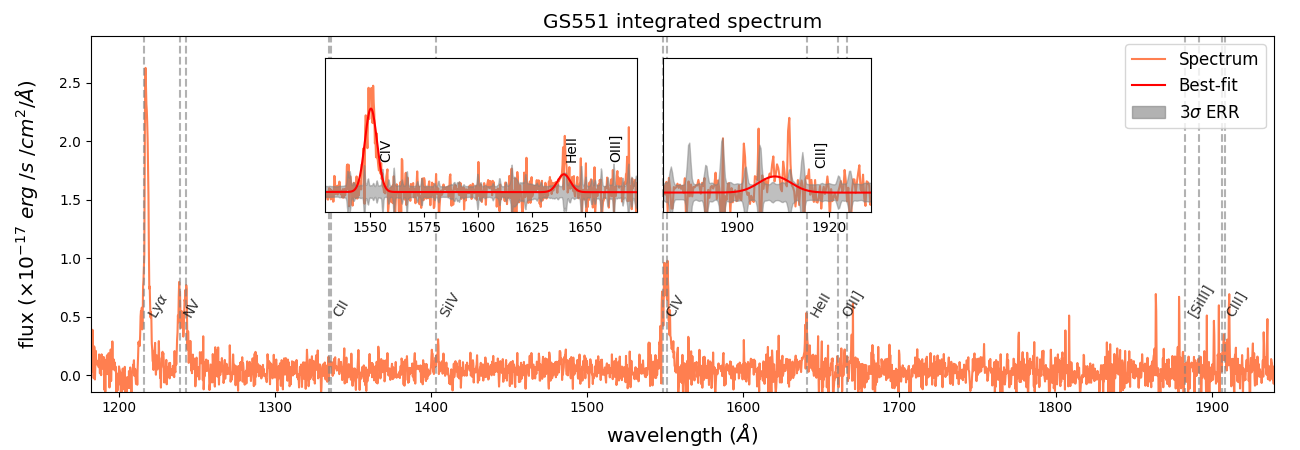}}

\caption{ MUSE spectra of GS551 (coral), showing AGN-ionised lines, according to UV diagnostics (Fig. \ref{fig:figUVBPT}). These spectra were obtained by integrating over the regions at S/N$~>4$ in Fig. \ref{fig:MUSEcutouts}; for visual purposes, the LAE2 spectrum was shifted, as labelled. The zoom-in insets show the same spectra in the vicinity of the lines used in the UV diagnostics, together with 1$\sigma$ errors for the LAE2 spectrum (green curve, mostly highlighting the position of strong sky line residuals).}\label{fig:GS551musespec}
\end{figure*}

We further examined the spectra of each \lya emitter to determine their ionisation sources. Of the six emitters, only LAE2 associated with GS10578 exhibits additional emission lines beyond \lya. We fitted the spectrum of LAE2, as well as those of the central X-ray AGN GS10578 and GS551, using Gaussian profiles (described in Sect. \ref{sec:analysis}). The best-fit models for these sources are displayed in Figs.\ref{fig:GS10578musespec} and \ref{fig:GS551musespec}; best-fit parameters are reported in Table~\ref{tab:fit_MUSE}.

High-ionisation lines such as 
C\,{\sc{iv}}$\lambda\lambda$1548,51 and  He\,{\sc{ii}}$\lambda$1640 lines are detected in the spectra of the two X-ray sources as well as in the LAE2 spectrum although at lower significance. 
In particular, LAE2 shows a strong \heii line detected at S/N$~\sim 20$; the \civ and \ciii lines are affected by sky line subtraction residuals, but their detections are associated with a reasonable significance ($> 5\sigma$).
The coronal line (CL) N\,{\sc{v}}$\lambda\lambda$1239,43 are observed only in the X-ray AGN, but its absence in LAE2 does not exclude AGN activity from this source  (e.g. \citealt{Maiolino2023a}). 
To explore the ionisation source in LAE2, we used the UV diagnostic diagram involving \civ/\ciii versus (\ciii + \civ)/\heii, as presented by \citet{Nakajima2022} and \citet{Feltre2016}. 

\begin{figure}[!htb]
\centering

\includegraphics[scale=0.62]{{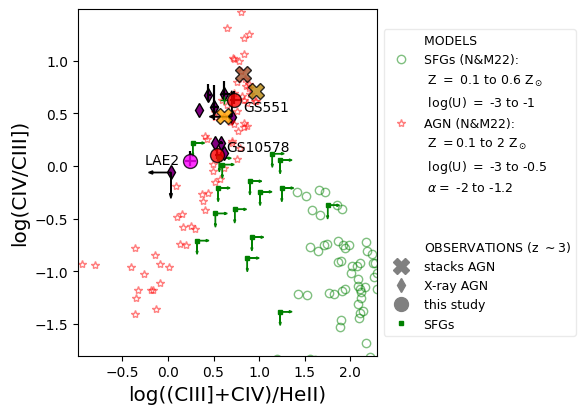}}

\caption{ UV diagnostic diagram. Small green and red empty symbols represent model predictions for SFGs and AGNs, respectively, from \cite{Nakajima2022}. See the legend for details about model parameters. 
Large circles refer to the MUSE measurements of the three sources in our sample, as labelled; MUSE resolution does not allow us to separate the contribution of primary and secondary AGNs, but GS551 and GS10578 line ratios are likely dominated by the primary (brightest) AGN. The figure also shows individual X-ray AGNs (diamonds) and measurements from stacked AGN spectra (crosses), as well as SFGs at $z\sim 3$ from the literature (green squares; see Sect. \ref{sec:resultsUV}). GS551, GS10578 and LAE2 occupy the AGN region of the diagram.
}\label{fig:figUVBPT}
\end{figure}

Figure~\ref{fig:figUVBPT} shows the positions of LAE2, GS10578, and GS551 in the UV diagnostic, together with nine additional X-ray AGNs at redshifts $z=2.6-4$ from other surveys. 
The flux ratios of five of them have been obtained from the analysis of publicly available VIMOS spectra from the VANDELS programme (\citealt{Mclure2018}), which provides some of the deepest spectra of intermediate redshift galaxies available to date. In particular, among the X-ray AGNs identified by \cite{Saxena2020} and \cite{Mascia2023} in VANDELS, we considered the obscured AGNs (i.e. without BLR emission in their UV spectra) with enough S/N to derive the flux ratios required for the UV diagnostic (i.e. excluding CDFS 019505). The fit analysis was performed using the same strategy presented for the MUSE data; the inferred flux ratios are reported in Table \ref{tab:fit_VIMOS}. 
In addition to the five AGNs from VANDELS, we included in Fig. \ref{fig:figUVBPT} the X-ray AGN UDS24561 at $z \sim 3$ discovered by \cite{Tang2022}, and the sources CDFS-057 ($z = 2.56$), CDFS-112a ($z = 2.94$), and CXO 52 ($z = 3.288$) from \cite{Dors2014}, using the flux ratios tabulated in the original papers. 
Figure \ref{fig:figUVBPT} also displays the flux ratios obtained from composite spectra of AGNs at $z\sim 3$ from \citet[][brown and light-brown crosses, indicating respectively their Class A and Class B AGN sources]{Alexandroff2013}, and \citet[][orange cross, from the stacked spectrum of seven X-ray obscured AGNs]{LeFevre2019}.

%\bigskip
%\bigskip

%\bigskip
%\bigskip

%\bigskip
%\bigskip

Non-active SFGs with strong high ionisation UV lines appear exceedingly rare in spectroscopic surveys (e.g. \citealt{Amorin2017, Mascia2023}). To represent them in Fig. \ref{fig:figUVBPT} we considered the catalogue of VANDELS sources in the UDS field by \cite{Talia2023}, and selected the sources for which at least one line is detected at S/N $> 3$ among \civ and \ciii (to derive at least upper or lower limit for the \civ/\ciii), and at least two lines are detected at S/N $> 3$ among  \ciii, \civ and \heii (to derive upper or lower limits for the (\civ + \ciii)/\heii).
These SFGs at $z\sim3$ are represented with green squares. 
We also plot the models of AGNs and SFGs as red stars and green circles, respectively, from \citet{Nakajima2022}. 
According to this UV diagnostic, LAE2 can be identified as an AGN emitter, sharing line ratios similar to those of GS10578, GS551, and other X-ray AGNs, as well of models of AGNs.

This UV diagnostic, therefore, confirms the presence of a third AGN in the GS10578 system, in addition to the AGN-A and AGN-B identified in the NIRSpec data. A more detailed discussion about the use of UV diagnostics is presented in Sect. \ref{sec:uvdiagnostics}.

\subsection{AGN bolometric luminosities}\label{sec:bolometricluminosities}

For all AGNs detected with NIRSpec, we computed the bolometric luminosities from the narrow \hb line, under the assumption that this emission comes from the NLR; bolometric corrections were computed using Eq. 3 of \citet{Netzer2019}, assuming a mean accretion disc inclination of $56^\circ$. These luminosities were preferred to the \oiii-based ones, because the latter are more dependent on the SMBH properties and the ionisation parameter (\citealt{Netzer2019}); for completeness, all \oiii (and \ha) luminosities are reported in Table \ref{tab:fit} for each target presented in this work. 

All \hb luminosities were corrected for extinction, using the measured Balmer decrements reported in Table \ref{tab:fit} and assuming a \cite{Cardelli1989} extinction law. For the AGN in LAE2 detected with MUSE, we computed the bolometric luminosity from an upper limit on the (undetected) continuum emission at 1400\AA, using  Eq. 3 by \cite{Netzer2019}, adopting the same accretion disc inclination mentioned above. We note that its \heii luminosity is comparable with that of GS10578-A; this might suggest that LAE2 and the primary AGN have comparable bolometric luminosities. 

All bolometric luminosities are reported in Table \ref{tab:DAGN}. As there is no rigorous way to estimate the uncertainties in the bolometric corrections introduced by \citet[][see their Sect. 2.4]{Netzer2019}, we preferred to not include error bars on the estimated bolometric luminosities; these should therefore be considered as order-of-magnitude estimates.

\subsection{Ionisation source}\label{sec:ionisationsource}

For two very close AGN pairs, GS551-A and -B with projected separation of 0.7 kpc, and GS10578-A and -B with separation of 4.7 kpc, we checked whether the primary AGN can power the emission in the companion galaxy. Specifically, we tested whether the line emission in GS551-B and GS10578-B can be explained by an external ionisation source (i.e. AGN-A), or requires an intrinsic source (i.e. a secondary AGN). %can  be due to intrinsic or external ionisation source.  

Following \cite{Keel2012}, we derived a lower limit to the incident ionising flux required for a line emitter taking into account the observed surface brightness in the recombination line \hb, and its distance from the primary AGN. No corrections for projection effects are considered, to obtain conservative estimates: a given companion will always lie farther from the nucleus of the primary AGN than our projected measurement, and hence require a higher incident flux.  
In a simple approximation, we considered the companion to be circular in cross-section as seen from the nucleus, so its solid angle is derived as $\theta = 2 \arctan (r/d)$, where $r$ is the radius of the emitting region and $d$ is the angular distance from the primary AGN. The required ionising luminosity is given from the observed quantities as

\begin{equation}
    L_{ion} > L(H\beta) \times k_{bol} \times \gamma \times  \left( \frac{4 \pi} {\theta^2} \right)
,\end{equation}
where L(\hb) is the \hb luminosity of the companion, $k_{bol}$ is the bolometric correction from \cite[][Sect. \ref{sec:bolometricluminosities}]{Netzer2019}, $\gamma = 0.14$ is the fraction of bolometric to ionising luminosity (considering the mean radio-quiet SED, following \citealt{Keel2019}), and $\theta$ is the subtended angle (in steradian).   
This required ionising luminosity has to be compared with the incident ionising luminosity coming from the primary AGN: L$_{inc} =$ L$_{bol} \times \gamma$ (following \citealt{Keel2019}). 

For GS551-B, we obtained a required L$_{ion} > 4\times 10^{45}$ \ergs, slightly higher than the incident L$_{inc} = 10^{45}$ \ergs. Although L$_{inc}$ and L$_{ion}$ are of the same order of magnitude, and we cannot definitely exclude the possibility of an external AGN ionisation, the latter scenario is very unlikely. In fact,  L$_{ion}$ has to be considered as a lower limit: assuming a $3\times$ larger physical (i.e. de-projected) distance, we would obtain $10\times $ higher L$_{ion}$, definitely incompatible with external AGN ionisation source. Therefore, our results are likely compatible with an intrinsic ionisation, hence with the presence of a secondary AGN.

For GS10578-B, we obtained a required L$_{ion}> 3\times 10^{45}$ \ergs, to be compared with the incident L$_{inc} \approx 6\times 10^{43}$ \ergs. In this case,  the ionising flux coming from the primary AGN is not enough to explain the emission in GS10578-B; we can therefore conclude that the latter is associated with a secondary AGN.

For the remaining sources, the required ionising luminosity is 10s$-10^{3}$ times  higher than the incident luminosity, definitely excluding the possibility of an external AGN ionisation source. Similar computations are not required for the pair LAE2 -- GS10578-A, as the two sources have comparable emission line luminosities in the UV (Table \ref{tab:fit_MUSE}); moreover, they are separated by 28~kpc.

\subsection{Shock ionisation}\label{sec:shocks}

We also tested the possibility that shocks could ionise the gas in the companions. Following \cite{Perna2020}, we used the shock model predictions of \ha luminosity from MAPPING V (\citealt{SutherlandDopita2017}), with magnetic field values in the range 10$^{-4}$--10 $\mu$G, shock velocities in the range 100--1000 \kms, pre-shock densities in the range 1--10 cm$^{-3}$ (to obtain \sii-based post-shock electron density in the range $50-4000$ cm$^{-3}$, in line with observations at high-$z$, e.g. \citealt{RodriguezDelPino2024,Lamperti2024}), and assuming a metallicity of 1 Z$_\odot$. These predicted luminosities are a factor of 4-to-2 dex smaller than our measurements (Table \ref{tab:fit}); therefore, we conclude that shock ionisation cannot be responsible for the emission in any of our secondary AGN sources.

\subsection{X-ray analysis of COS1638}\label{sec:xray}

X-rays  have proven inefficient in detecting dual AGNs at $z>1$ due to the limited sensitivity and angular resolution of current X-ray facilities (\citealt{Sandoval2023}).
Nevertheless, here we investigate the X-ray morphology of one of our dual AGN system, covered with \chandra observations. Additional information about X-ray emission of all other systems is presented in Appendix~\ref{sec:Atargets}.

\begin{figure*}[!htb]
\centering
\includegraphics[width=0.8\linewidth, trim=0mm 8mm 0mm 10mm]{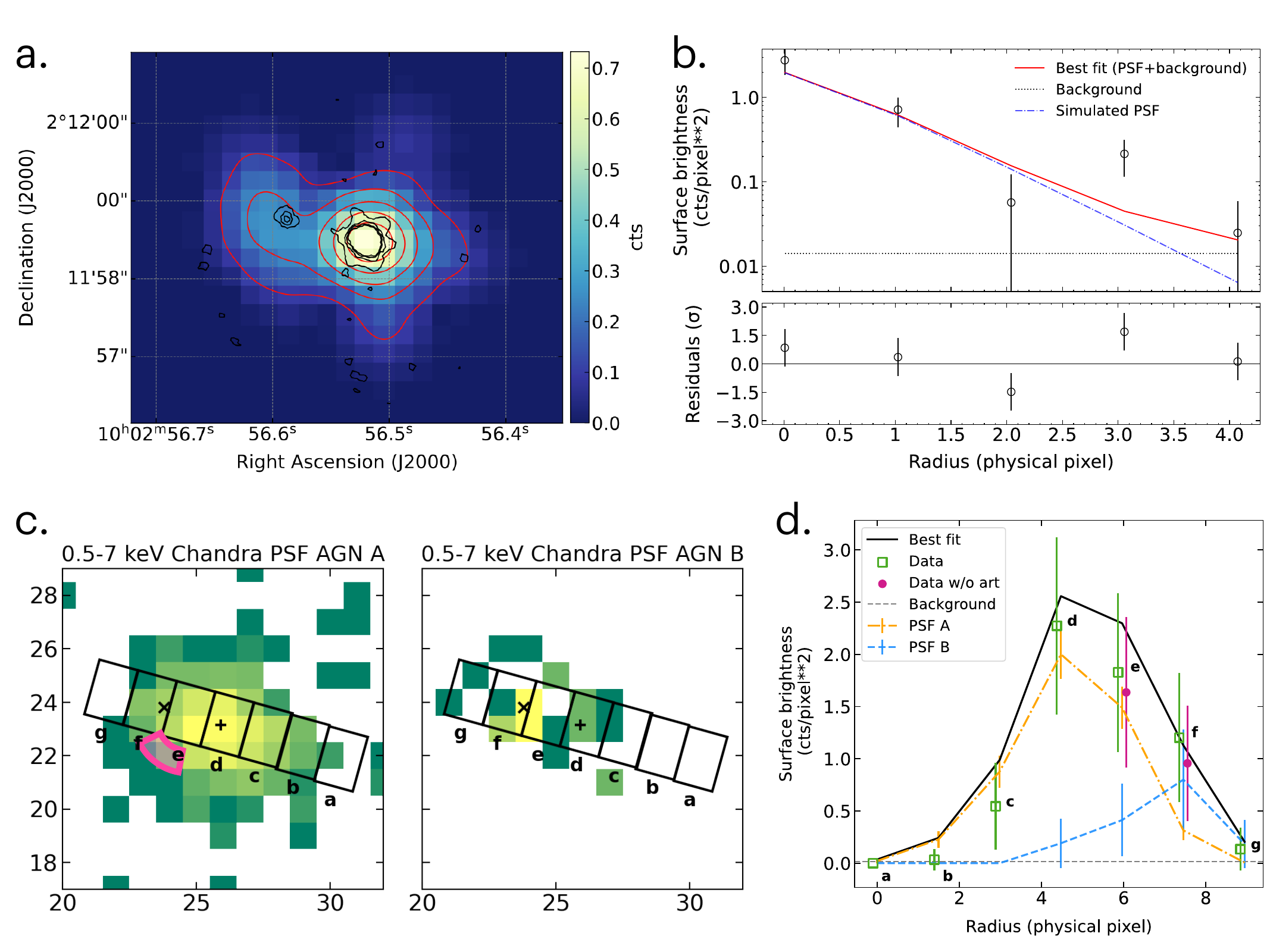}

{\phantomsubcaption\label{fig:Xray.a}
 \phantomsubcaption\label{fig:Xray.b}
 \phantomsubcaption\label{fig:Xray.c}
 \phantomsubcaption\label{fig:Xray.d}
}

\caption{ COS1638 X-ray emission. (\subref{fig:Xray.a}): \chandra image (obsID 15253) in the 0.5-7 keV observed-frame energy band, smoothed and sub-binned (1/2 pixel binning, i.e. 0.25\arcsec pixel scale) for visualisation purposes. NIRSpec \oiii and X-ray contours are overlayed in black and in red, respectively. (\subref{fig:Xray.b}): radial profile of COS1638 in obsID 15253 extracted from the native-pixel event file, after masking the pixels affected by the PSF hook, using a set of 5 annuli ($\delta r \simeq 0.5$\arcsec) centred on the peak of the source emission (which corresponds to COS1638-A). The radial profile was fit with the radial PSF model (see Sect.~\ref{sec:xray}) plus a constant set to the measured background level (model: PSF+constant). The red solid line shows the best fit, the blue dash-dotted line shows the simulated PSF profile, while the black dotted line shows the measured background level. The observed radial profile shows hints ($\sim$2$\sigma$) of an X-ray excess at the distance of COS1638-B ($\sim$2-3 native Chandra pixels, corresponding to $\sim1.2$\arcsec). (\subref{fig:Xray.c}): 0.5-7 keV images of the simulated PSFs at the positions of COS1638-A (marked with a `+') and COS1638-B (`x'). The lettered boxes show the extraction regions of the radial profiles in panel \subref{fig:Xray.d}. The magenta region shows the position of the hook feature, whose corresponding pixels were masked in the \chandra data before extracting the radial profiles.
(\subref{fig:Xray.d}): 0.5--7 keV (observed energy range) radial profiles of COS1638 extracted from ObsID 15253, from masked (magenta) and unmasked (green) data. The black solid line marks the total X-ray profile obtained from fitting the data with the two PSFs (dash-dotted and blue dashed lines for COS1638-A and COS1638-B, respectively) and the background (grey dashed line). 
Thus, the X-ray morphology overall hints at the presence of a second peak at the position of COS1638-B.
}\label{fig:Xray}
\end{figure*}

We inspected the X-ray morphology of the dual AGNs associated with COS1638 through the \chandra exposures of the COSMOS-\textit{Legacy} survey. Given the survey strategy \citep{elvis2009, civano2016}, the target is placed at different distances from the \chandra aim point in each observation (each with same exposure time), with an off-axis angle of at least $\theta\sim3.5'$. As a result, the single-exposure point spread function (PSF) at the position of the primary AGN is 
of at least $\sim1.2$\arcsec. 
Figure \ref{fig:Xray.a} shows the 0.5--7 keV (observed-frame, sub-binned and smoothed) \chandra image obtained in the best conditions (obsID 15253, 50~ks, $\theta\sim3.5'$, PSF$~\sim$~1.2\arcsec), with overlayed \oiii contours. 

Thus, we investigated the X-ray radial profile of COS1638 in the \chandra exposure obsID 15253, using the Chandra Ray Tracer (ChaRT v.2\footnote{\url{https://cxc.cfa.harvard.edu/ciao/PSFs/chart2/runchart.html}}) online tool (see Fig. \ref{fig:Xray} for details). 
We simulated the \chandra PSFs at the positions of COS1638-A and COS1638-B with ChaRT, feeding the ray-tracing code with the respective target and aim-point coordinates, the exposure time and the respective 0.5--7 keV source spectrum. The source spectrum extracted from this observation is integrated over the two AGN components and dominated by COS1638-A and, thus, we used its best fit for COS1638-A. For COS1638-B, we built an obscured powerlaw model ($N_{\rm H}=5\times10^{23}~\rm cm^{-2}$), given that it is a type 2 AGN, setting the X-ray flux as equal to that obtained converting its \oiii luminosity through the relation of \cite{Lamastra2009}. 
We ran 50 ChaRT simulations for both PSFs, each of which was then projected on the detector plane using MARX v.5.5.2 \citep{davis2012_marx}. We merged the output of MARX with the task {\tt dmmerge} in CIAO v4.15 \citep{fruscione2006} to best reproduce both \chandra PSFs. We show the 0.5-7 keV maps of the two PSFs in Fig. \ref{fig:Xray.c}. 

It is known that the PSF of both cameras on board \chandra presents a hook-like feature dependent on the observation characteristics (\citealt{Koss2015}) that ChaRT and MARX do not account for. We thus used the {\tt make\_psf\_asymmetry\_region} task of CIAO to uncover the pixels affected by such an asymmetry and mask them in the observation event file. We show the asymmetry region overlayed to the PSF of COS1638-A for clarity. 
We then extracted the radial profile of the entire system, from the event file masked for the PSF asymmetry, by using a set of 5 annuli ($\delta r \simeq 1$pix $\simeq 0.5$\arcsec) centred on the peak of the source emission (which corresponds to COS1638-A) and fitted it with the radial PSF model (obtained from the PSF of COS1638-A) plus a constant set to the measured background level (model: PSF A+constant), using Sherpa 4.15.1 \citep{sherpa2023}. Background contribution was estimated from a 15\arcsec-radius source-free circle in the proximity of our target. 
Figure \ref{fig:Xray.b} shows the 0.5--7 keV (observed-frame) X-ray profile from the data and the best fit from Sherpa. The observed profile shows hints ($\sim$2$\sigma$) of an X-ray excess at the distance of COS1638-B ($\sim$2-3 native Chandra pixels, corresponding to $\sim1.2$\arcsec). 
We also repeated this exercise restricting the extraction areas to those shown in Fig. \ref{fig:Xray.c}, to reduce noise contamination. We show in Fig. \ref{fig:Xray.d} the radial profile from the \chandra data, both masked and unmasked for the PSF asymmetry, the total radial profile of the two PSFs and the contribution of each PSF. 
Hence, despite the degraded PSF due to the off-axis position, the X-ray morphology suggests the presence of a second peak at the position of COS1638-B.

We tested the possibility of this peak being due to high SF activity, rather than to a secondary AGN; the measured SFR$~\simeq$~1200~\Msunyr (see Appendix.~\ref{sec:Atargets}) translates to L[0.5--8~keV]$~\simeq8\times10^{42}$~\ergs (\citealt{Mineo2014}). Thus, we addressed the detectability of a starburst galaxy with such a high SFR by assuming a reasonable spectral shape for the X-ray binary emission \cite[and references therein]{yang2020_xcigale}. The expected X-ray flux of such starburst galaxy would be almost one order of magnitude lower than the flux limit of COSMOS-\textit{Legacy} and subsequently undetectable in obsID 15253 alone. Thus, the X-ray emission of COS1638 supports the presence of a secondary AGN in COS1638-B.

\begin{table*}[h]
\centering
\caption{Dual (and triple) AGNs.}\label{tab:DAGN}%
\begin{tabular}{@{}|l|c|c|ccc|c|c|c|c|c@{}}
\hline
target & redshift  & RA \& DEC & \multicolumn{3}{c|}{Separation} & log(M$_*$)& log(L$_{\rm bol}$) &log(L$_{\rm X}$) & C$_{AGN}$
 \\
  &    &  (degrees) & (\arcsec)  & (kpc) & (\kms) & (M$_\odot$)& (\ergs)& (\ergs) & \\
  {\footnotesize (1)} & {\footnotesize (2)}  & {\footnotesize (3)}  & {\footnotesize (4)}  &{\footnotesize (5)}  &{\footnotesize (6)}  &{\footnotesize (7)}  &{\footnotesize (8)} &
  {\footnotesize (9)}&{\footnotesize (10)} \\
\hline
COS1638-A & $3.5079$ & 150.735557 +2.19953 & -- & -- & -- & $10.9\pm 0.4$ & $46.7$ & 44.5 & 2,3,4\\
COS1638-B & $3.5109$  & 150.735847 +2.19962 & 1.1  & 8.2  & 200 & $11.1\pm 0.4$& $46.2$ & -- & 1,2,4$^\ddagger$ \\
\hline 
GS551-A    & $3.7034+$   & 53.124385 $-$27.85169  & $-$ & $-$& -- &$10.9\pm 0.3$ &$46.0$  &44.4 & 1,2,4$^*$,5\\
GS551-B$^\diamond$   & $3.7022+$  & 53.124274 $-$27.85166  &0.4 & 2.9 & $-25$ & -- &$44.5$  &--& 1\\
\hline 
Eastfield (-A)$^\natural$   & $3.7149$   & 150.061668 +2.37887  &  $-$ &$-$ & -- & $9.5 \pm 0.2$& $44.8$ & -- &1,2$^\ddagger$ \\
Mr. West (-B)$^\natural$   & $3.7096$   & 150.061348 +2.37877 & 1.2 & 8.8 & $-340$ & $9.4\pm 0.3$ & $\lesssim44.3$ & -- &1,2$^\ddagger$\\
\hline 
COS1656-A & $3.5101$  & 150.271546 +1.61383 & $-$ & $-$ &-- & $11.0\pm 0.2$ &$45.8$ &44.4 & 1,2,4$^*$\\
COS1656-B & $3.5084$+  & 150.271589 +1.61424 & 1.4 & 10.4 & $-60$ & $9.7\pm 0.3$ &$44.3$ & -- &1\\
\hline 
GS10578-A  & $3.0647$+   & 53.165325 $-$27.81415  & $-$ & $-$ &-- & $11.2\pm 0.2$ & $44.6$ & 44.6&1,2,4$^*$,5\\
GS10578-B  & $3.0644$+   & 53.165532 $-$27.81415  & $0.7$ &  4.7 &$-30$& --&$44.2$& --& 1 \\
LAE2    & $3.0674$   & 53.164688 $-$27.81489  & 3.6 &  28 & $200$ & $8.3\pm 0.4$ &$\lesssim 44.6$ &-- & 5\\
\hline
\end{tabular}
\tablefoot{
Column (1): target name. Eastfield and Mr West identified the newly discovered companion of Jekyll \& Hyde, following \cite{Perez-Gonzalez2024}. Column (2): Redshift, as measured from the narrower Gaussian component in the UV and optical spectra (typical uncertainty of 0.0002). Column (3): Coordinates RA \& DEC. Columns (4), (5) and (6): projected separation and velocity offset of the secondary (and tertiary) AGN from the primary nucleus. Column (7): stellar mass, as inferred from spectral energy distribution analysis (see Sect. \ref{sec:stellarmasses}); for COS1638-A and its close companion we provided order of magnitude estimates assuming a conservative $M_{dust}$/M$_*$ (see Sect. \ref{sec:stellarmasses}). Column (8): bolometric luminosity, measured from the NLR \hb flux using Eq. 3 of \citet[][typical uncertainty of 0.1 dex, not taking into account the bolometric correction scatter]{Netzer2019}, for all but the AGN in LAE2 for which UV continuum is instead used (see Sect. \ref{sec:bolometricluminosities}).  Column (9): absorption-corrected X-ray luminosity; see Appendix \ref{sec:Atargets}. Column (10): flags on specific criteria used to identify AGN: 1= BPT diagram, 2= prominent \oiii outflow, 3= BLR emission, 4= X-ray emission, 5= UV line ratios of \cite{Nakajima2022} and \cite{Feltre2016}. 

$^\diamond$: Candidate secondary AGN, as incident ionising flux from the primary AGN could be responsible for the fluxes in this target (see Sect. \ref{sec:ionisationsource}).\\
$^\natural$: Candidate AGN, as the presence of highly ionised gas might also be explained by a fading and/or highly obscured AGN in Hyde (\citealt{Perez-Gonzalez2024}), even though no evidence of AGN emission is found in Hyde.\\
$^\ddagger$: tentative evidence; see Appendix \ref{sec:Atargets}. \\
$^*$: X-ray emission likely associated with both sources; see Appendix \ref{sec:Atargets}. 
}
\end{table*}

\subsection{Stellar masses of AGN host galaxies}\label{sec:stellarmasses}

To infer the stellar masses of the host galaxies of the multiple AGNs presented in this work, we exploited their multi-wavelength information.

To study the physical properties of the stellar populations in Eastfield and Mr West (Fig.~\ref{fig:jekyllspec}), we analysed the prism (0.6--5.3$\mu$m) NIRSpec IFS observations  in a spaxel-by-spaxel basis following the method described in \citet{PerezGonzalez2023}, and adapted to the modelling of NIRSpec IFS data by \cite{DEugenio2024NatAs}. These data were reduced following the same procedure described in Sect. \ref{sec:observations}. 
Based on this method, we derived integrated stellar masses of $\sim 10^9$ M$_\odot$ for both AGN host galaxies, by adding all the spaxels associated with these systems. In particular, for Mr West we considered a circular region with $r = 0.2$\arcsec centred at the position of the nucleus (AGN-B in Fig. \ref{fig:jekyllspec}), while for Eastfield we considered a larger ($r=0.3$\arcsec) region to encompass its ring-shaped structure (reddish region in Fig. \ref{fig:jekyllspec}).  
More detailed analysis of the physical properties of the stellar populations in this very complex system is presented in \cite{Perez-Gonzalez2024}; here we stress that the mass measurements depend only slightly on the chosen apertures.

For all other systems observed with NIRSpec IFS, we analysed through spectral energy distribution (SED) fitting the COSMOS (GOODS-S) UV-to-NIR multiwavelength photometry collected by \citet{Weaver2022} and \citet{Merlin2021}, and the far-IR data by \cite{Jin2018} and \cite{Shirley2021}. We also included in the analysis the ALMA sub-millimetre detections (for COS1638) and 3$\sigma$ upper limits (for all other sources). For the \lya emitter in the vicinity of GS10578, associated with the AGN-C (Fig. \ref{fig:lyaoiiimaps}, top left), we used publicly available NIRCam and HST measurements collected by \cite{Rieke2023}.

We modelled the SED with the Code Investigating GALaxy Emission (CIGALE v0.11.0, \citealt{Boquien2019}), a publicly available state-of-the-art galaxy SED-fitting software. This code disentangles the AGN contribution from the emission of the host by adopting a multicomponent fitting approach and includes attenuated stellar emission, dust emission heated by star formation, AGN emission (both primary accretion disc emission and dust-heated emission), and nebular emission (see \citealt{Circosta2018, Circosta2021} for further details).
The more detailed SED analysis will be presented in Circosta et al., in prep. Here we report that the primary AGNs are hosted in massive ($\approx 10^{11}$ M$_\odot$) galaxies; the secondary and tertiary AGN hosts have smaller masses, in the range $10^8-10^{10}$ M$_\odot$ (see Table \ref{tab:DAGN}). Unfortunately, no M$_*$ measurements can be obtained for COS1638-B, GS551-B and GS10578-B, as they are not resolved as individual sources in the above mentioned catalogues (they are too close to the primary AGN): indeed, the photometry used for the primary AGN host may be contaminated by these companions; as a result, the estimates we obtain may include the contribution of both components.

A separate handling is required for the type 1 COS1638-A. Stellar mass estimates in the range $10^{8}-10^{12}$ M$_\odot$ are reported in the literature for the host galaxy of the primary AGN (e.g. \citealt{Jin2018,Weaver2022}); this divergence is likely due to the  non-thermal AGN emission dominating the optical continuum (Circosta et al, in prep.), which is precluding an SED-based estimate of M$_*$. However, an order of magnitude estimate of the stellar mass can be obtained considering our M$_{BH} = (4\pm 1)\times 10^8$ M$_\odot$ measurement (see Appendix \ref{sec:Atargets}) and assuming the local black hole mass-stellar mass relation (\citealt{Kormendy2013}), M$_* \sim 10^{11}$ M$_\odot$. A further independent estimate can be obtained for both primary and secondary AGN hosts from the far-IR-based dust mass M$_{dust} \sim 10^{11}$ M$_\odot$ reported in \cite{Liu2019} and referring to the dual AGN system (because of the poor spatial resolution of far-IR data). Archival ALMA data (2016.1.00463.S, PI: Y. Matsuda) tracing far-IR emission at $\sim 1$ mm shows that COS1638-B is brighter than the primary AGN host (62\% of the emission comes from the AGN-B host; see Appendix \ref{sec:Atargets}). Assuming that 62\% of the dust mass is due to the AGN-B host, and considering a conservative M$_{dust}$/M$_* = 0.02\pm 0.01$ (\citealt{Donevski2020}) we obtained log(M$_*$/M$_\odot$) $\sim 10.9\pm 0.4$, for COS1638-A, and $\sim 11.1\pm 0.4$, for COS1638-B. 
We note that the far-IR properties of this system, as well as its X-ray properties and optical emission profiles and line ratios appear to be similar to the BR1202--0725 (\citealt{Zamora2024}) dual AGN systems.

\section{Discussion}\label{sec:discussion}

\subsection{ AGN classification based on optical diagnostics at $z > 2.6$}\label{sec:opticaldiagnostics}

Several approaches have been proposed to test the presence of an AGN from the study of rest-frame optical spectra. Among them, the use of optical diagnostic diagrams such as the BPT (\citealt{Baldwin1981}, see our Fig. \ref{fig:BPT}), and the detection of forbidden high ionisation lines, known as CLs, with ionisation potentials IP $> 54.4$ eV (the \heii edge; e.g. \citealt{Oliva1994, Moorwood1997}). In this section we discuss their possible use and reliability for the AGN classification of the newly discovered sources COS1638-B, GS551-B, Eastfield, Mr. West, COS1656-B, and GS10578-B.

Usually, \nev $\lambda 3435$  and \fevii $\lambda 6087$ are the most prominent  forbidden CLs in the optical range. Unfortunately, \nev is not covered by our NIRSpec observations. 
The \fevii line, which is the brightest CL in the $3500-7500\AA$ range, is usually only 1-10\% of the strength of \oiii (\citealt{Murayama1998, Mingozzi2019}). 
Moreover, quantitative studies on the prevalence of CLs, even in the local Universe, remain limited. CLs are thought to be more common in type 1 AGNs (e.g. \citealt{Lamperti2017, Cerqueira-Campos2021}). Interestingly, however, CLs remain undetected in near- and mid-infrared wavelength regions of the well-known local type 1 AGNs observed with JWST, Mrk 231 (Ulivi et al., in prep.; \citealt{AlonsoHerrero2024}). This underscores the challenges in relying on CLs for AGN classification (see also \citealt{Perna2024arp}).

The \nev  and \fevii high ionisation lines are not detected in our newly discovered AGN sources, nor in their composite spectrum (\fevii/\oiii $\lesssim 3$\%; see Fig. \ref{fig:nonXraycomposite}). This is likely because of their type 2 nature; in fact, a composite spectrum of six X-ray detected, type 2 AGNs in our sample similarly shows no clear evidence of CLs although they are much brighter than the secondary AGN (Fig. \ref{fig:Xraycomposite}). 
However, we stress that the prevalence of CLs in the AGN population is unconstrained yet, at any redshift. 
Therefore, other diagnostics are considered in this work to identify AGN emission in our sources.  

The BPT diagnostic provides an unambiguous classification for the newly identified sources in Figs. \ref{fig:cos1638spec}--\ref{fig:gs551spec}, as their flux ratios fall into the Seyfert region of the diagram (see Fig. \ref{fig:BPT}). This area is deemed free from contamination by SFGs, according to the latest observational and theoretical findings. 
On the one hand, the contaminating processes that may mimic AGN-like signatures, that is the presence of post-AGB stars (e.g. \citealt{Wylezalek2018}), can be excluded thanks to the very high \oiii/\hb and \ha equivalent width (EW). In particular, the measured EW(\ha) of $15 - 150~\AA$ (see Table \ref{tab:fit}) are significantly higher than those expected for post-AGB stars ($< 3 \AA$, e.g. \citealt{Belfiore2016}). 
On the other hand, hot young stars cannot be responsible for Seyfert-like line ratios, according to both theoretical predictions (e.g.  \citealt{Topping2020, Runco2021, Nakajima2022}) and observations: recent JWST observations proved that non-active galaxies at $z > 2.6$ can have high \oiii/\hb but relatively low \nii/\ha with respect to local Seyfert and active galaxies at high-$z$ (e.g. \citealt{Sanders2023, Nakajima2023, Sun2023}). The BPT diagnostic in Fig. \ref{fig:BPT} shows a compilation of SFGs at $z = 2.6 - 7$, from the CEERS (\citealt{Calabro2023}) and JADES (\citealt{Scholtz2023}) surveys. These sources avoid the Seyfert region of the BPT, and tend to occupy the same locus as SFGs at cosmic noon (light-olive shaded area in the figure). On the contrary, high-z  AGN can populate both the Seyfert region and the locus of non-active galaxies, as proved by our measurements from GA-NIFS (red circles, associated with X-ray detected AGNss and broad-line AGNs) and from JADES measurements (orange points, classified as AGNs on the basis of additional diagnostics, from \citealt{Scholtz2023}; see also e.g. \citealt{Harikane2023, Kocevski2023, Maiolino2023c, Duan2024}).

Summarising, AGNs at $3\lesssim z \lesssim 6$ can have {\it i)} optical line ratios very similar to those of local Seyfert and $z \sim 2$ active galaxies, or {\it ii)} line ratios similar to SFGs at $z>1$. The former, with log(\nii/\ha) $\gtrsim -0.5$ and log(\oiii/\hb) $\gtrsim 0.75$, are more likely hosted in massive systems (see Table \ref{tab:fit}). The latter, with log(\nii/\ha) $\lesssim -1$ and log(\oiii/\hb) $\gtrsim 0.75$, are likely hosted in galaxies with  smaller masses (and lower metallicity, e.g. \citealt{Scholtz2023}). This dichotomy aligns with predictions from photoionisation models developed over the last decades (see e.g. \citealt{Dors2024} and references therein).
Most importantly, the Seyfert region of the BPT diagram is free from contamination by SFGs. All arguments collected in this section reinforce the BPT-based AGN classification for the newly identified sources shown in Figs. \ref{fig:cos1638spec}--\ref{fig:gs551spec}.

\subsection{ AGN classification based on UV diagnostics at $z > 2.6$}\label{sec:uvdiagnostics}

Direct observations of high-ionisation lines in the UV regime, such as N\,{\sc{v}}$\lambda\lambda$1239,43, C\,{\sc{iv}}$\lambda\lambda$1548,51, He\,{\sc{ii}}$\lambda$1640, also in combination with other UV transitions, serve as robust indicators of AGN activity in distant galaxies (e.g. \citealt{Mascia2023, Scholtz2023, Maiolino2023b}). 
In this section we focus on the use of \nv, a very high ionisation line  (IP $\sim 80$ eV), 
and the UV diagnostic diagram \civ/\ciii versus (\ciii + \civ)/\heii (Fig. \ref{fig:figUVBPT}) 
to distinguish between AGN and SF ionisation in high-$z$ sources, and confirm the presence of an AGN in LAE2.

As noted by \cite{Maiolino2023a}, the absence of \nv  cannot definitively rule out AGN activity, as its intensity relative to other UV lines varies significantly. 
For instance, \nv is a few times stronger than \heii in the X-ray detected AGN GS551 (see Fig.~\ref{fig:GS551musespec}, Table \ref{tab:fit_MUSE}), but it is not detected in other X-ray AGNs reported in the literature (see e.g. \citealt{Law2018, Tang2022}). 
The ratio  \nv/\civ can vary from 0.01 to 1 in type 1 AGNs (see Fig. 6 in \citealt{Maiolino2023a}), and is often $< 1$ in obscured AGNs at $z > 2$ (e.g. \citealt{Silva2020, Mignoli2019}; see also Tables \ref{tab:fit_MUSE}, \ref{tab:fit_VIMOS}). Consequently, the non-detection of \nv in the MUSE spectrum of LAE2 does not exclude the presence of an accreting SMBH in this source.

To test the reliability of the UV diagnostic diagram \civ/\ciii versus (\ciii + \civ)/\heii presented in Fig. \ref{fig:figUVBPT},
we incorporated measurements for nine X-ray detected AGNs at $z \sim 2.6 - 4$ and flux ratios from composite spectra of AGNs at $z\sim 3$ reported in the literature. By including these additional measurements, we ensured a robust comparison between UV line ratios observed in LAE2 and those of confirmed AGNs. 
The flux ratios of these bona fide AGNs align closely with those of LAE2, as well as with predictions for AGN ionisation models by \citet[][see also \citealt{Gutkin2016}]{Nakajima2022}. 

In contrast, SFGs with strong high-ionisation UV lines appear exceedingly rare in spectroscopic surveys (e.g. \citealt{Amorin2017, Mascia2023}). 
To display their population in Fig. \ref{fig:figUVBPT}, we considered the catalogue of VANDELS sources in the UDS field by \cite{Talia2023}, and selected galaxies where the detected UV lines, or a combination of detections and upper limits, allowed their inclusion in the diagnostic plot. 
Figure \ref{fig:figUVBPT} shows that all SFGs have upper limits that point towards the locus of \cite{Nakajima2022} predictions for SF systems (see also \citealt{Feltre2016}). 
Most of the $\sim 900$ sources in the \cite{Talia2023} VANDELS catalogue cannot be reported in this diagnostic diagram, as the UV emission lines are usually very faint or undetected even with $> 20$ hours integration time (see also \citealt{Mascia2023}). 

Well detected \civ, \ciii and \heii have been reported in the literature for a few non-active galaxies (e.g. \citealt{LeFevre2019,Mascia2023}); their flux ratios cover the bottom-right part of the UV diagram in Fig. \ref{fig:figUVBPT}, as shown by \citet[][see their Fig. 11]{Nakajima2018}. Moreover, a handful of sources classified as local dwarf galaxies (\citealt{Berg2016} and \citealt{Senchyna2017}) and $z \sim 2$ compact SF galaxies (\citealt{Amorin2017}) can reside in the AGN region of the UV diagnostic (see Figs. 11 and 14 in \citealt{Nakajima2018}). 
Nevertheless, these peculiar sources can typically be distinguished from AGNs by their strong low-ionisation lines 
such as [Si\,{\sc{iii}}]$\lambda\lambda$1883,92 and O\,{\sc{iii}}]$\lambda\lambda$1661,66 (e.g. \oiiiuv/\civ  $> 1$; see Fig. 4 in \citealt{Senchyna2017}, Fig 2 in \citealt{Berg2016}, Supplementary Fig. 4 in \citealt{Amorin2017}). Such low ionisation lines are  usually faint or undetected in obscured AGNs at high-z (and in LAE2; see Tables \ref{tab:fit_MUSE} and \ref{tab:fit_VIMOS}; see also \citealt{Talia2017}), consistent with theoretical predictions (e.g. \citealt{Nakajima2022}). In tables \ref{tab:fit_MUSE} and \ref{tab:fit_VIMOS} we reported additional UV line ratios usually taken into account in alternative UV diagrams commonly adopted in the literature (see e.g. \citealt{Mignoli2019, Mingozzi2024}); also these alternative UV diagnostics show similarities between LAE2 and the X-ray sources collected in this work, and therefore confirm the AGN classification for LAE2.    

In conclusion, LAE2 is the only source undetected in X-ray sitting at the position of X-ray AGNs at similar redshifts (from this study, and from  \citealt{Dors2014, Saxena2020, Tang2022, Mascia2023, LeFevre2019}); all SFGs from VANDELS are instead below the location of X-ray AGNs. LAE2 also differs from local dwarf galaxies and compact SFGs at the cosmic noon in terms of \oiiiuv/\heii. 
All evidence, including UV line diagnostics, comparisons to X-ray AGNs, and the absence of strong low-ionisation features, strongly supports the classification of LAE2 as an AGN.

\subsection{Dual AGN fraction at $z\sim 3$}\label{sec:dualagnfraction}

The fraction of dual AGN systems in the early Universe provides crucial insights into galaxy mergers and the growth of SMBHs. Moreover, AGN pairs are pivotal for forecasting gravitational wave background levels and event rates in PTA experiments and for the future LISA mission.
Cosmological simulations, such as those homogenised by \citet{DeRosa2019} and, more recently, by \citet{PuertoSanchez2024}, predict dual AGN fractions of
$1\%-6$\% for physical separations d$_{\rm 3D}<30$ kpc
and luminosities $L_{\rm bol}> 10^{43}$ \ergs in the redshift range $z\sim 1-6$ (see also \citealt{Steinborn2016, Rosas-Guevara2019, Chen2023}). 
\cite{Volonteri2022} reported an expected fraction $\sim 4\%$ for dual AGNs at $z\sim 3$ with slightly stricter criteria of d$_{\rm 3D}<30$ kpc and $L_{bol}> 10^{44}$ \ergs.
However, to align the threshold criteria  of these theoretical works with the GA-NIFS observations, we should consider projected separations d$_{\rm p}<10$~kpc and even higher bolometric luminosities (see Table \ref{tab:DAGN}). In fact, all theoretical works show that predictions are sensitive to the threshold criteria used to define the parent population and the number of duals, in addition to the specific dynamical and physical implementations and numerical methods.

\begin{figure}[!t]
\centering

\includegraphics[scale=0.455]{{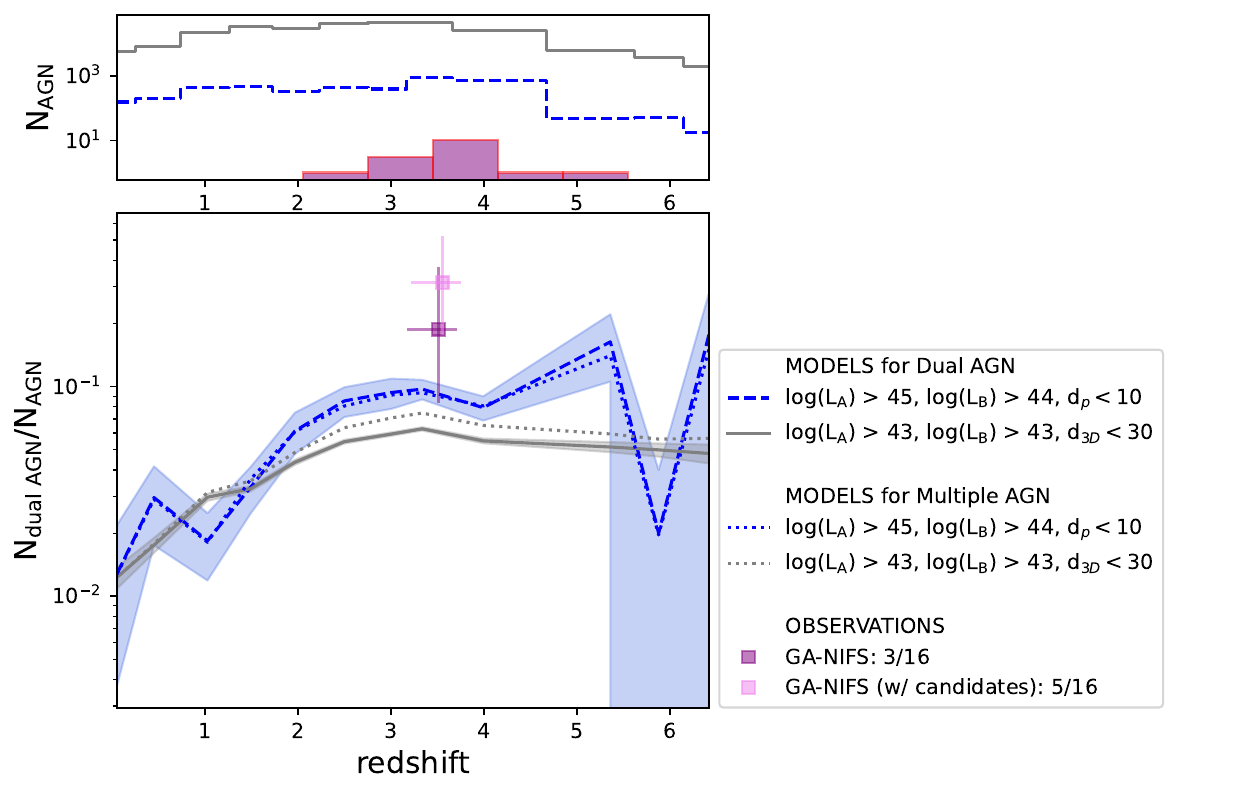}}

\caption{Dual AGN fraction as a function of redshift. The purple square shows the inferred dual AGN fraction from GA-NIFS observations, for the three newly discovered dual AGNs; instead, the violet symbol displays the fraction obtained also considering the two candidate dual AGNs;
both measurements include 1$\sigma$ Poisson errors for small numbers. The different curves represent the predicted AGN fractions obtained from an extension of the \citet{Volonteri2022} simulations up to $z\sim 6$, passing different threshold criteria (as labelled). Notably, the figure shows that the GA-NIFS measurements moderately exceed the predictions obtained for thresholds that mimic our observational criteria, hence with projected distances d$_{\rm p}< 10$~kpc, log(L$_{\rm bol, \ A}$) $> 45$ (for the primary AGN and the parent population) and log(L$_{\rm bol, \ B}$) $> 44$ (for the secondary AGN), shown with a blue dashed curve (and Poissonian 1$\sigma$ intervals in lightblue). 
The top panel shows the parent sample distributions of GA-NIFS sources (purple histogram) and simulated AGNs (blue for log(L$_{\rm bol, \ A}$) $> 45$, grey for log(L$_{\rm bol, \ A}$) $> 43$); we note that simulations at $z>5$ suffer from poor statistics. 
}\label{fig:AGNfraction}
\end{figure}

To ensure the match between GA-NIFS observations and predictions, we presented in Fig. \ref{fig:AGNfraction} the AGNs extracted from the Horizon-AGN simulation, using an identical approach to those presented in \cite{Volonteri2022}, but extended to higher redshifts ($z \sim 6$) and considering dual AGNs with projected separations d$_{\rm p}<10$~kpc and $<30$~kpc and different AGN luminosities. 
The newly derived fractions at $z\sim 2-5$ for the most luminous dual AGNs are slightly higher compared to earlier reports in the literature (e.g. \citealt{DeRosa2019}). This discrepancy can be attributed to three key factors: 
(i) the fraction increases if the secondary is allowed to be fainter than the primary (see also the blue curve in Fig. 16 of \citealt{Volonteri2022}), (ii) the most luminous AGNs are more likely hosted in massive galaxies, which, at high redshift, are often found in over-dense regions and hence more likely to have close active companions, and (iii) no cuts were imposed on the mass of the SMBHs, something that is often done in cosmological simulations to improve the purity of the sample, but at the cost of decreasing completeness (e.g. \citealt{PuertoSanchez2024}). Further details regarding the assumptions considered for deriving the predictions reported in the figure are discussed in Appendix \ref{sec:Ahorizon}.

Figure \ref{fig:AGNfraction} also shows the fraction of dual AGNs derived from our JWST observations, of $\sim 20-30$\%. Specifically, the lower value is inferred considering only the three confirmed dual AGNs systems out of the 16 AGNs in  GA-NIFS, hence excluding the GS551 and Jekyll systems; the higher value is instead derived also considering these two candidate dual AGN systems. Our inferred fractions are moderately above predictions ($\sim 10$\%) from cosmological simulations that mimic our observational criteria, that are:  
d$_{\rm p}<10$~kpc, log(L$_{\rm bol, \ A}$) $> 45$ (for the primary AGN and the parent population) and log(L$_{\rm bol, \ B}$) $> 44$ (for the secondary AGN), shown with a blue dashed curve Fig. \ref{fig:AGNfraction}. 
As expected, our measurements significantly exceed the predicted fractions obtained considering  log(L$_{\rm bol, \ A}$) and log(L$_{\rm bol, \ B}$) $> 43$, and d$_{\rm 3D}< 30$~kpc (grey curve), consistent with previous works (e.g. \citealt{DeRosa2019, Rosas-Guevara2019, Volonteri2022}).

We also considered the predictions for dual and multiple AGN systems with d$_{\rm 3D}< 30$~kpc (black dotted curve) and d$_{\rm p}< 10$~kpc (blue dotted curve), taking into account that 
i) for GS10578 and GS551 we could investigate the presence of additional active companions at larger distances from the primary AGN thanks to MUSE observations, and that
ii) we actually discovered the GS10578 triple AGN system with projected separations $d_{\rm p} = 4.7$~kpc (AGN-A, B) and $d_{\rm p} = 28$~kpc (AGN-A, C). 
Also in these cases, Horizon-AGN simulations provide fractions $\lesssim 10$\% at $z\sim 3$.

%\bigskip
%\bigskip

It is important to note that different simulations employ varying models for galaxy and black hole physics, leading to discrepancies in predicted AGN and dual AGN populations. These differences include the initial mass of the BH seeds, seeding conditions in galaxies, BH accretion, AGN and stellar feedback, as well as the implementation of dynamical friction from gas, stars, and dark matter particles. Additionally, simulation resolution, volume, and subgrid prescriptions influence the predictions (e.g. \citealt{Habouzit2021}). The evolution of the dual AGN fractions over time also depends on the single and dual AGN selection criteria (AGN luminosity, SMBH mass, host stellar mass). All these factors complicate direct comparisons with observations. 
\citet{PuertoSanchez2024} examined various simulations (Illustris, \citealt{Genel2014, Vogelsberger2014}; TNG50, TNG100, TNG300, \citealt{Pillepich2018, Nelson2019}; Horizon-AGN, \citealt{Dubois2016, Volonteri2016}; EAGLE, \citealt{Schaye2014, Rosas-Guevara2019}; SIMBA, \citealt{Dave2019,Thomas2019}; BlueTides, \citealt{Feng2016}; and Astrid, \citealt{Ni2022, Bird2022}), highlighting the impact of different modelling choices on AGN demographics. The simulations analysed provide dual AGN fractions between 1 and 6 \% at $z\sim 3$ for AGNs with L$_{\rm bol,\ A, \ B}$ $> 10^{43}$~\ergs at physical distances $d_{\rm 3D}\leq 30$~kpc (see their Fig. 2). These fractions remain nearly unchanged for luminosities $> 10^{44}$~\ergs (see their Fig. B1).  
Unfortunately, their results are not directly applicable to GA-NIFS due to the different luminosity thresholds (and lack of SMBH and host stellar mass estimates) in our sample.
Nevertheless, \citet{PuertoSanchez2024} showed that predicted dual AGN fractions remain relatively stable across simulations (within a factor of $\sim 2$ at $z\sim 3$), unlike the absolute number densities of single and dual AGNs, which can vary by a few orders of magnitude. In practice, while the overall AGN population has a stronger dependence on how AGNs are modelled in the simulations, the dual AGN fraction is is less dependent on specific choices. For instance, the SIMBA simulation (\citealt{Dave2019,Thomas2019}) models BH accretion inspired by loss of angular momentum through torques, whereas Horizon-AGN adopts the Bondi-Hoyle-Lyttleton formalism. The two simulations also adopt different BH seeding methods (i.e. one using a fixed stellar mass threshold, and the other is based on local stellar density); yet their predicted dual AGN fractions differ by less than a factor of 2 (see Figs. 2 and B1 in \citealt{PuertoSanchez2024}). This suggests that the comparison presented in Fig. \ref{fig:AGNfraction} between observations and predictions could be broadly valid across different simulation frameworks, albeit with some inherent uncertainties.

In summary, our results point to the presence of a large population of dual AGNs in the early Universe, up to $\sim 20-30$\% of the total number of active galaxies. This fraction is moderately above predictions ($\sim 10$\%) from cosmological simulations that mimic our observational criteria (blue dashed curve in Fig. \ref{fig:AGNfraction}).  However, the limited number of observed systems (two dual AGNs and two candidate dual AGNs within $d_{\rm p} \lesssim 10$~kpc, and a triple AGN within $d_{\rm p} < 30$~kpc, out of 16 AGN systems) precludes us from conclusively determining whether Horizon-AGN cosmological simulations underestimate the number of dual AGNs at high-$z$. 
Moreover, while our comparison with Horizon-AGN provides valuable insights, it is important to acknowledge that the predicted dual AGN fraction can vary slightly across different cosmological simulations. A comprehensive study of these variations would require a detailed analysis of each simulation's specific parameters and implementations, which is beyond the scope of this work.

\subsection{Towards a possible revised understanding of AGNs at high-z}

Our results suggest a substantial population of dual AGNs in the early Universe, with fractions reaching $\sim 20\%-30\%$, moderately exceeding our newly extracted predictions from the Horizon-AGN cosmological simulations (
$\sim10\%$) tailored to our observational criteria. Although the limited GA-NIFS sample size precludes definitive conclusions about whether simulations systematically underestimate dual AGN fractions, we note that our findings suggest that multiple AGN systems are more prevalent than previously identified in earlier observational efforts (e.g. \citealt{Silverman2020, Shen2023, Sandoval2023}).

Recent observational studies further support this view, suggesting that multiple AGN systems may indeed be more frequent than previously thought  (e.g. \citealt{Lemon2022, Yue2023, Maiolino2023c, Spingola2019, Matsuoka2024, Pensabene2024,Perna2024gs133, Travascio2024}). In particular, \cite{Spingola2019} and \cite{Schwartz2021} found evidence for a sub-kiloparsec dual AGN associated with a lensed quasar at $z = 3.28$; these authors provided an independent, indirect measurement of the fraction of dual AGNs in the early Universe, of $\sim 16$\%, in line with our results. Moreover, \cite{Ubler2024zs7} reported the discovery of a candidate dual AGN at $z \sim 7$, observed as part of a different GA-NIFS sub-sample of 11 galaxies at $z > 6$. 
In another compelling case, \citet{Zamora2024} discovered an AGN pair in the overdense region surrounding BR1202--0725 at $z=4.7$. This system consists of a sub-millimetre galaxy (SMG)-QSO pair with projected separation of 24~kpc, and multiple line emitters (e.g. \citealt{Carniani2013, Drake2020}). Using NIRSpec IFS and \chandra data, \citet{Zamora2024} revealed a dusty SMBH in the SMG. Therefore, BR1202--0725 underscores the growing evidence for multiple accreting SMBHs in high-density environments (see also e.g. \citealt{Coogan2018,Perna2023}).

Lastly, a recent study by \cite{Li2024} identified 28 bona fide dual AGN in a sample of 78 X-ray-selected AGNs at $0.5 < z < 4.5$, using HST and JWST imaging from the COSMOS-Web survey (\citealt{Casey2023})\footnote{ Our dual AGN COS1638 and COS1656 are not covered by COSMOS-Web. In the Eastfield and Mr. West group, NIRCam barely detects the brightest object, Jekyll. Regarding the remaining GA-NIFS targets in the COSMOS field, all but COS1118 appear as isolated sources in NIRCam images; the bright source at $\sim 1.8$\arcsec north-east from COS1118 is classified as an M star on the basis of its NIRSpec spectrum.}. Although these results require spectroscopic confirmation, they are well aligned with our findings (see e.g. their Fig. 8).

This growing body of evidence suggests a revision in our understanding of AGN populations at high redshift, with implications for models of SMBH growth, galaxy mergers, and the emergence of structure in the early Universe.

\section{Conclusions}\label{sec:conclusions}

This study leverages JWST/NIRSpec integral field observations to investigate the prevalence of dual AGNs at $z\sim 3$, analysing a sample of 16 AGNs observed as part of the JWST GA-NIFS survey. Using the high spatial and spectral resolution of NIRSpec, we searched for companion line emitters within a field of $\sim 20\times 20$~kpc$^2$ around each AGN.
For two targets, archival VLT/MUSE data were incorporated, extending the search area to projected separations of up to $\sim 50$~kpc. The main results of this work are summarised below.

We detected line-emitting companions near the bright AGNs in nine of the 16 GA-NIFS targets. Among these, we identified five systems hosting multiple AGNs, including one triple AGN and four dual AGNs (two considered candidates). The projected separations range from 3 to 28~kpc. These systems were primarily classified using classical BPT optical emission-line diagnostics diagrams from JWST/NIRSpec spectra. 
Both primary and secondary AGNs in these systems show \oiii/\hb and \nii/\ha line ratios that are incompatible with those of distant star-forming galaxies; instead, they  are  similar to those of AGNs and Seyferts at any redshifts.
The secondary AGNs show high \nii/\ha line ratios, indicative of metal-enriched environments, consistent with the massive ($\sim 10^{11}$~M$_\odot$) host galaxies of the primary AGN inferred via SED fitting. These companions have estimated stellar masses of log(M$_*$/M$_\odot$)~$\gtrsim 9.4$. Moreover, they show ionisation sources consistent with intrinsic AGN activity rather than star formation, further supporting their classification as secondary AGNs.

Our AGN classifications were further supplemented by additional arguments. 
The third AGN in the triple system, discovered via MUSE UV observations, was classified as an active galaxy based on UV line ratios typical of AGNs, rather than SFGs at similar redshifts. This third AGN is also observed in rest-frame optical photometry (e.g. HST, JWST/NIRCam) and tentatively detected in the far-IR (with ALMA; see \citealt{Scholtz2024gs10578}; Circosta et al., in prep.); it is associated with a low stellar mass ($\sim 10^8$~M$_\odot$). 
Another notable system, COS1638, displays a secondary AGN with exceptionally fast outflows, characteristic of high-luminosity AGN ($>10^{46}$~\ergs), pronounced far-IR brightness compared to AGN-A, and tentative ($3\sigma$) X-ray detection consistent with AGN activity. These features underscore the utility of multi-wavelength data in strengthening AGN classifications.

This study more than doubles the known population of multiple AGN systems at $z\gtrsim 3$ with separations under 30 kpc (e.g. \citealt{Mannucci2023, PuertoSanchez2024}). 
The serendipitous discovery of multiple AGNs in $\sim 20\%-30\%$ of the systems in our sample exceeds the predictions from cosmological simulations commonly reported in the literature. 
We extracted dual AGN fractions from Horizon-AGN cosmological simulations mimicking the GA-NIFS observational criteria in terms of separation and $L_{\rm AGN}$, deriving an expected fraction $\lesssim 10\%$ at $z\sim 3$. Therefore, our observational results moderately exceed these predictions.

The findings of this study have important implications for our understanding of the prevalence of dual AGNs in the early universe and their role in the co-evolution of galaxies and supermassive black holes. The high number of dual AGNs identified in the GA-NIFS survey may indicate that cosmological simulations might underestimate the incidence of dual AGNs at high redshift. However, the current sample size used in this work is limited to 16 sources (12 within the narrower range $3\lesssim z \lesssim 4$ where dual AGNs were identified) and a larger sample is necessary to confirm this conclusion.

The redshift range of $3\lesssim z \lesssim 4$ is a key period for studying SMBH interactions and mergers. Simulations indicate that the population of dual AGN peaks before the general AGN population (see Fig. 3 in \citealt{PuertoSanchez2024}); moreover, mergers of SMBHs require long timescales ($\sim1$~Gyr, \citealt{Tremmel2017}). Therefore, dual AGNs at $3\lesssim z \lesssim 4$ represent a key population as they are likely connected with the peak of AGN activity and the rates of coalescing SMBHs, and, consequently, gravitational wave events at cosmic noon.

This study shows that JWST/NIRSpec IFS is exceptionally well-suited to identifying dual AGNs in the distant universe thanks to its high sensitivity and (spatial and spectral) resolution.
This work demonstrates the power of JWST in uncovering hidden AGN populations and motivates further research to refine estimates of the dual AGN fraction across cosmic time. Continued observations with JWST and other advanced telescopes will provide invaluable insights into the formation, evolution, and impact of these intriguing systems.

\begin{acknowledgements}

We thank the anonymous referees for their constructive feedback that helped to improve the quality of this work.
We are grateful to C. Vignali and E. Bronzini for useful discussion on the processing of \chandra data. We thank K. Nakajima for providing the theoretical model grids published by \cite{Nakajima2022}. We thank I. Delvecchio, C. Spingola, C. Lemon, M. Annunziatella, M. Villar Martin, L. Costantin, A. Feltre, and M. Yue for helpful comments on an earlier version of this manuscript.

MP, SA, BRP, and IL acknowledge support from the research project PID2021-127718NB-I00 of the Spanish Ministry of Science and Innovation/State Agency of Research MCIN/AEI/ 10.13039/501100011033. 
MP acknowledges support through the grant RYC2023-044853-I, funded by  MICIU/AEI/ 10.13039/501100011033 and FSE+.
IL acknowledges support from grant PRIN-MUR 2020ACSP5K\_002 financed by European Union – Next Generation EU.
RM acknowledges support by the Science and Technology Facilities Council (STFC), by the ERC Advanced Grant 695671 ``QUENCH'', and by the UKRI Frontier Research grant RISEandFALL; RM is further supported by a research professorship from the Royal Society.
AJB acknowledges funding from the ``First Galaxies'' Advanced Grant from the European Research Council (ERC) under the European Union’s Horizon 2020 research and innovation programme (Grant agreement No. 789056).
MV acknowledges funding from the French National Research Agency (grant ANR-21-CE31-0026, project MBH\_waves) and from CNES for the space mission LISA.
MP, GC and EB acknowledge the support of the INAF Large Grant 2022 "The metal circle: a new sharp view of the baryon cycle up to Cosmic Dawn with the latest generation IFU facilities". 
H\"U acknowledges support through the ERC Starting Grant 101164796 ``APEX''.
%
%We thank K. Nakajima for providing the theoretical model grids published by \cite{Nakajima2022}.

\end{acknowledgements}

%
%-------------------------------------------------------------
%               Appendices have to be placed at the end                                  

%-------------------------------------------------------------
%\begin{thebibliography}
%\end{thebibliography}

\bibliographystyle{aa}
\bibliography{aa53430-24corr.bib}

\begin{appendix}
\onecolumn

\section{Observation logs and fit results}

We report the information about NIRSpec observations in Table \ref{tab:log}; Tables \ref{tab:fit}, \ref{tab:fit_MUSE}, and \ref{tab:fit_VIMOS} outline the NIRSpec, MUSE, and VIMOS spectral fit results, respectively.

\begin{table*}[h]
\centering

\tabcolsep 3.5pt 
\caption{JWST/NIRSpec and VLT/MUSE observations used in this work.}\label{tab:log}%
\begin{tabular}{@{}lccccccc@{}}
\hline
target & $z$ & RA DEC & instrument & configuration & programme ID  & Obs. date & Exposure (s) \\
\hline
COS2949 & 2.048& 150.4029167 1.87889 & NIRSpec &  G235H/F170LP & 1217  & 23-04-2023  & 3560  \\

COS1638$^\star$ & 3.508 & 150.7355417 2.19956 & NIRSpec & G235H/F170LP  & 1217  & 26-04-2023  & 3560  \\

COS349 & 3.509 & 150.0043771 2.03890 & NIRSpec & G235H/F170LP  & 1217  & 09-05-2023  & 3560  \\

COS1656$^\star$& 3.510 & 150.2715833 1.61385 & NIRSpec  & G235H/F170LP & 1217  & 21-04-2023  & 3560  \\
%HZ10   & & 150.2471250 1.55534 & NIRSpec   & 1217  & 22-05-2023  & 18207 \\

COS590 & 3.524 & 149.7554125 2.73853 & NIRSpec & G235H/F170LP  & 1217  & 09-05-2023  & 3560  \\

COS1118 & 3.643& 149.8791917 2.22584 & NIRSpec & G235H/F170LP  & 1217  & 23-04-2023  & 3560  \\

Jekyll$^\star$& 3.715 & 150.0614600 2.37868 & NIRSpec & G235H/F170LP & 1217  & 01-05-2023  & 15872 \\
\hline

GS10578$^\star$& 3.065 & 53.1653058 -27.81413 & NIRSpec & G235H/F170LP  & 1216  & 12-08-2022  & 14706  \\
                &    &                  & MUSE & WFM  & 1101.A-0127       &  mosaic\footnotemark[1] &  340500 \\

GS19293 & 3.117 &53.0539287 -27.74771 & NIRSpec  & G235H/F170LP & 1216  & 06-02-2023  & 14706   \\

GS20936 & 3.246 &53.1178958 -27.73438 & NIRSpec  & G235H/F170LP & 1216  & 26-08-2022  & 14706   \\

GS811  & 3.466 & 53.1846596 -27.88097 & NIRSpec & G235H/F170LP  & 1216  & 12-09-2022  & 3560 \\

GS133  & 3.472 & 53.0206229 -27.74223 & NIRSpec & G235H/F170LP  & 1216  & 12-09-2022  & 3560 \\

GS774  & 3.585 & 53.1744017 -27.86740 & NIRSpec & G235H/F170LP  & 1216  & 12-09-2022  & 3560   \\

GS551$^\star$ & 3.703 & 53.1243458 -27.85163 & NIRSpec & G235H/F170LP  & 1216  & 12-09-2022  & 3560 \\
                &          &            & MUSE  & WFM & 094.A-0205  & 25-12-2014  &  3600 \\

GS539 & 4.755 & 53.1220783 -27.93878 & NIRSpec & G235H/F170LP & 1216  & 12-09-2022  & 3501 \\
 &   &   & NIRSpec &  G395H/F290LP & 1216  & 12-09-2022  & 3501 \\
GS3073 & 5.550 & 53.0788554 -27.88416 & NIRSpec & G395H/F290LP  & 1216  & 25-09-2022  & 18207  \\

\hline
\end{tabular}
\tablefoot{The multiple AGN systems presented in this work are marked with $^\star$ symbols.
GS10578 MUSE-UDF cube was obtained from a collection of different exposures, covering three semesters.
}
\end{table*}

\begin{table*}[h]
\centering
\tabcolsep 3.5pt 
\caption{NIRSpec fit results.}\label{tab:fit}%
\begin{tabular}{@{}lcccccccc@{}}
\hline
Target & $\sigma$ & Log(\oiii /\hb) & Log(\nii /\ha) & \ha/\hb & {EW(\ha)}& Log(L(\oiii)) & Log(L(\ha)) \\
      & (\kms) &  &  &  &  $(\AA)$&(\ergs) & (\ergs) \\
\hline
COS1638-A & $1200_{-90}^{+45}$  &   $0.08_{-0.03}^{+0.08}$  &  $-0.45_{-0.55}^{+0.51}$   &   $6.4_{-0.5}^{+2.1}$ & $40_{-10}^{+30}$ & $42.6$ & $42.9$\\
COS1638-B & $1030_{-36}^{+120}$  &  $0.60_{-0.10}^{+0.20}$   &  $0.40_{-0.30}^{+0.20}$  &  $9.1_{-2.5}^{+8.0}$ & $100_{-11}^{+50}$&  $41.5$ & $41.3$ \\

\hline
COS1656-A & $666_{-6}^{+10}$  & $1.01_{-0.02}^{+0.03}$ &  $0.15_{-0.01}^{+0.02}$   &    $4.2_{-0.2}^{+0.3}$& $63_{-1}^{+2}$ &$42.3$ & $41.9$   \\
COS1656-B & $127\pm7$  &  $0.60_{-0.03}^{+0.04}$   & $-0.32_{-0.02}^{+0.03}$   & $2.9_{-0.2}^{+0.4}$  & $85_{-3}^{+5}$& $41.4$ & $41.4$ \\
Clump & $83\pm 4$ & $0.89\pm 0.03$ & $-0.95_{-0.06}^{+0.09}$ & $3.5_{-0.1}^{+0.3}$ & $183_{-6}^{+8}$ & 41.8 & 41.4\\

\hline
Eastfield & $180_{-30}^{+10}$ & $0.86_{-0.04}^{+0.06}$ & $-0.35_{-0.04}^{+0.09}$ & $2.5_{-0.2}^{+0.4}$ & $165_{-9}^{+7}$ & $42.1$ & $41.8$\\
Mr. West & $275\pm40$ & $0.88_{-0.11}^{0.08}$ & $0.06_{-0.05}^{+0.10}$ & $3.4_{-0.7}^{+1.1}$ &$62_{-4}^{+8}$ &$41.8$ & $41.5$\\
Clump & $85\pm 2$ & $0.93_{-0.02}^{+0.05}$ & $-0.68_{-0.08}^{+0.02}$ & $3.6_{-0.2}^{+0.5}$ & $127\pm 3$ &  41.8 & 41.4\\

\hline
GS551-A & $450_{-5}^{+10}$ & $1.06_{-0.02}^{+0.04}$ & $-0.26_{-0.04}^{+0.10}$ & $3.3_{-0.1}^{+0.4}$ & $150_{-5}^{+4}$& $43.1$ & $42.5$ \\
GS551-B & $55_{-2}^{+3}$ & $1.10_{0.02}^{+0.03}$ & $-0.56_{-0.04}^{+0.06}$ & $3.2_{-0.2}^{+0.3}$ & $>38$ & $41.9$ & $41.3$\\

\hline
GS10578-A & $730\pm 10$& $1.01_{-0.02}^{+0.12}$ & $0.88_{-0.09}^{+0.07}$& $3.2_{-0.5}^{+0.7}$ & $165_{-1}^{+1}$ &$42.2$ & $41.5$\\ 
GS10578-B & $114\pm 5$  & $0.97_{0.05}^{+0.13}$ & $-0.73_{-0.02}^{+0.06}$ & $3.3_{-0.3}^{+0.1}$  & $72_{-5}^{+4}$ &$41.5$   &  $41.1$   \\
Clump & $190_{-7}^{+8}$ & $0.97_{-0.06}^{+0.09}$ & $<-0.78$ & $4.2_{-0.4}^{+1.3}$ & $55_{-2}^{+3}$ & 41.2 & 40.9\\

\hline
\end{tabular}
\tablefoot{ Best-fit spectroscopic results for the most prominent emission lines detected with NIRSpec. The velocity dispersion $\sigma$ refers to the total \oiii emission line profile.  \oiii and \ha luminosities have typical uncertainty of 0.1 dex, and are not corrected for extinction. COS1638-A flux ratios  are affected by degeneracy between NLR (systemic and outflow) and BLR emission.
}
\label{tab:nirspeclineratios}
\end{table*}

%\begin{landscape}
\begin{table*}[h]
\tabcolsep 5pt 
\caption{MUSE fit results.}\label{tab:fit_MUSE}%
\begin{tabular}{@{}lccccccccccc@{}}
\hline
target & $dv$ & $\sigma$ &   $\frac{\rm CIII]}{\rm HeII}$ &    $\frac{\rm CIV} {\rm HeII}$ &   $\frac{\rm CIV}{\rm CIII]}$ &   $\frac{\rm CIII] + CIV}{\rm HeII}$ &$\frac{\rm NV}{\rm HeII}$ &$\frac{\rm CIV}{\rm NV}$ &$\frac{\rm OIII]}{\rm HeII}$ &   L(\heii)  \\
     & (\kms)& (\kms) &  &   &  & & & & & (\ergs)  \\
\hline
%\hline 
GS10578 & $-260_{-40}^{+50}$ & $445\pm 30$ & $0.15_{-0.03}^{+0.04}$ & $0.24_{-0.02}^{+0.04}$ & $0.11_{-0.04}^{+0.05}$ & $0.54_{-0.02}^{+0.04}$ & $0.19_{-0.04}^{+0.09}$& $0.05_{-0.04}^{+0.09}$&$<-1.22$ &$41.2\pm 0.1$\\
LAE2 & $208_{-4}^{+10}$ & $120_{-9}^{+24}$ & $-0.08\pm 0.06$ & $0.03_{-0.05}^{+0.07}$ & $0.05_{-0.05}^{+0.09}$ & $0.24_{-0.05}^{+0.07}$ & $< -0.74$ & $>0.33$ &$<-0.85$& $40.8\pm 0.2$ \\
\hline
GS551 & $-2_{-30}^{+10}$ & $540\pm 20$ & $0.02\pm 0.09$ & $0.63_{-0.04}^{+0.08}$ & $0.63_{-0.05}^{+0.09}$ & $0.73_{-0.05}^{+0.07}$ & $0.78_{-0.05}^{+0.06}$& $-0.14_{-0.02}^{+0.03}$ & $<-0.44$ &$41.9\pm 0.1$\\
\hline
\end{tabular}
\tablefoot{Best-fit spectroscopic results for the most prominent emission lines detected with MUSE. $dv$ is the velocity offset with respect to the optical (NIRSpec) zero-velocities in the primary AGN. Luminosities and flux ratios are in log-scale.}
%\footnotetext[1]{}
\end{table*}

\begin{table*}[h]
\tabcolsep 5pt 

\caption{VIMOS fit results.}\label{tab:fit_VIMOS}%
\begin{tabular}{@{}lccccccccccc@{}}
\hline
target & z & $\sigma$ &   $\frac{\rm CIII]}{\rm HeII}$ &    $\frac{\rm CIV} {\rm HeII}$ &   $\frac{\rm CIV}{\rm CIII]}$ &   $\frac{\rm CIII] + CIV}{\rm HeII}$ &$\frac{\rm NV}{\rm HeII}$ &$\frac{\rm CIV}{\rm NV}$ &$\frac{\rm OIII]}{\rm HeII}$ &   L(\heii)  \\
     &  & (\kms) &  &   &  &  & & & & (\ergs)  \\
\hline
%\hline 

UDS020721 & 2.519 & $270_{-30}^{+90}$ & $-0.22_{-0.11}^{+0.15}$ & $<-0.33$ & $-0.11_{-0.11}^{+0.05}$ & $<0.01$ & -- & -- & $<-0.60$ & $41.9\pm 0.1$\\

CDFS005827 & 3.034 & $400_{-25}^{+60}$ & $0.06_{-0.03}^{0.10}$ & $0.38_{-0.02}^{+0.03}$ & $0.31_{-0.02}^{+0.12}$ & $0.55_{-0.02}^{+0.04}$ &$-0.01_{-0.04}^{+0.1}$ & $0.39_{-0.05}^{+0.09}$ & $-0.51_{-0.06}^{0.23}$ & $41.9_{-0.2}^{+0.1}$\\

CDFS019824 & 3.476 & $310_{-10}^{+50}$ & $-0.24_{-0.04}^{0.09}$ & $0.31_{-0.03}^{+0.08}$ & $0.56_{-0.09}^{+0.03}$ & $0.42_{-0.02}^{+0.08}$ & $0.10_{-0.08}^{+0.07}$ & $0.24_{-0.04}^{+0.07}$& $< -0.95$ & $42.7\pm 0.1$\\

UDS025482 & 3.526 & $240_{-50}^{+90}$ & $-0.01_{-0.04}^{+0.03}$ & $0.53_{-0.04}^{+0.06}$ & $0.56_{-0.20}^{+0.10}$ & $0.65_{0.04}^{+0.11}$ & $0.06_{-0.06}^{+0.09}$ & $0.47_{-0.04}^{+0.08}$& $<-0.32$ & $41.7_{-0.1}^{+0.1}$\\

UDS018960 & 3.941 & $250_{-25}^{+8}$ & $-0.20_{-0.15}^{+0.18}$ & $0.40_{-0.03}^{+0.06}$ & $0.60_{-0.11}^{+0.18}$ & $0.50_{-0.04}^{+0.08}$ & $-0.07_{-0.03}^{+0.06}$ & $0.48_{-0.03}^{+0.04}$& $-0.37_{-0.08}^{+0.11}$ & $42.5_{-0.1}^{+0.1}$\\

\hline
\end{tabular}
\tablefoot{Best-fit spectroscopic results obtained from the analysis of available X-ray detected AGNs from the VANDELS survey. Note that the UDS020721 spectrum does not cover the \nv emission line. CDFS019824 corresponds to GS133, also presented in \citet{Perna2024gs133}.}
%\footnotetext[1]{}
\end{table*}

\section{Individual targets}\label{sec:Atargets}

Here we report a brief description of the main properties of each system studied in this work.

\subsection{COS1638 (dual AGN)}

COS1638 hosts a bright type 1 AGN with an X-ray luminosity log($L_{2-10\ \rm keV}$/\ergs) $\sim 44.5$ (\citealt{Marchesi2016}). Its optical spectrum (Fig. \ref{fig:cos1638spec}) shows prominent BLR emission in both \ha and \hb lines, and strong iron emission in the vicinity of the \oiii lines (see also \citealt{Trakhtenbrot2016}). From the best-fit described in Sect.~\ref{sec:analysis}, we derived M$_{BH} = (4\pm 1)\times 10^8$ M$_\odot$, a BLR \hb-based bolometric luminosity log(L$_{bol}$ / \ergs) $= 46.5\pm 0.1$ (using the Eqs. 38 and 25 by \citealt{Dallabonta2020}), and hence an Eddington ratio of $0.8\pm 0.2$. These values have been obtained from the BLR \hb luminosity (corrected for extinction, E(B--V) = 0.6 from the BLR Balmer decrement, assuming a \cite{Cardelli1989} extinction law), consistent with that inferred from the NLR and reported in Table \ref{tab:DAGN}.

The secondary AGN (COS1638-B) shows very high \nii/\ha and \oiii/\hb line ratios (Table \ref{tab:fit}), and extremely broad profiles with $\sigma$(\oiii) $\sim$ 1100 \kms (Fig. \ref{fig:cos1638spec}). Such broad profiles cannot be due to starburst driven outflows (e.g. \citealt{Cicone2016}); instead they are explained by the presence of energetic AGN outflows (e.g. \citealt{Villar2020}). Narrower \oiii emission is also detected between the primary and secondary AGNs, indicating gravitational interactions (Fig. \ref{fig:cos1638spec}, left panel). The complex morphology of this system, with different clumps at several kpc from the primary AGN host, is also observed with HST/ACS images (e.g. tracing rest-frame UV continuum). By analysing archival ALMA data (2016.1.00463.S, PI: Y. Matsuda) we found that COS1638-B is brighter than the primary AGN host at $\sim 1$ mm (obs-frame): we measured a total integrated flux of 11.4 mJy, with 62\%  of which associated with COS1638-B. Therefore, a significant amount of dust is located at the position of the secondary AGN, in line with the extreme Balmer decrement we measure in NIRSpec data (\ha/\hb $\sim 9$). 
Taking into account their relative ALMA continuum fluxes, and the CIGALE best-fit results, we also inferred the star-formation rate (SFR) of $\approx 1200$ \Msunyr for AGN-B, and $\approx 800$ \Msunyr for AGN-A. 
We also inspected the X-ray morphology of this dual AGN through the \chandra exposures of the COSMOS-\textit{Legacy} survey, which supports the presence of a secondary AGN in COS1638-B.

\subsection{GS551 (candidate dual AGN)}

GS551 is a massive ($\sim 10^{11}$ M$_{\odot}$) galaxy hosting an obscured AGN.
The NIRSpec IFS observations of GS551 are shown in Fig. \ref{fig:gs551spec}: we identified two bright \oiii emitters, one at the nuclear position of the central galaxy, dubbed GS551-A, and one at $\sim 0.4$\arcsec north-west, dubbed GS551-B (also detected in HST-ACS and WFC3 images, as well as in JWST/NIRCam and MIRI images; Circosta et al., in prep.). In addition, diffuse \oiii emission at $\sim 5$ kpc on the east side of GS551 is detected. 

The GS551-A spectrum (Fig. \ref{fig:gs551spec}) shows a strong outflow and AGN emission line ratios, according to the standard BPT diagram (Fig. \ref{fig:BPT}). The second source similarly shows AGN line ratios, with $\sim 10$ times fainter features likely powered by intrinsic ionisation of an obscured AGN (see Sect.~\ref{sec:ionisationsource}). 

The 7MS \chandra observations do not allow for the identification of the dual AGNs, as their spatial separation is lower than the \chandra resolution at the position of the target; the (absorption-corrected, $N_{\rm H}=9\pm1\times10^{23}$ cm$^{-2}$) X-ray luminosity log($L_{0.5-7\ \rm keV}$/\ergs) $= 44.35\pm0.05$ reported by \cite{liu2017_cdfs} may therefore be attributed to both sources.

The characterisation of the GS551 \lya nebula has been presented in \cite{Herenz2019}, \cite{denBrok2020}, and \cite{Schmidt2021}. Here we highlight that GS551 shows a very complex \lya halo on scales of $\sim 15$ kpc; interestingly, we also detected two additional \lya structures $\sim 20$ kpc south-east and $\sim 40$ kpc north-west of the central galaxy, and at similar redshifts, as shown in Fig. \ref{fig:lyaoiiimaps}. 
Furthermore, an even more distant \lya emitter was identified by \cite{denBrok2020}, located  $\sim 80$~kpc to the north-east of GS551 and also at a similar redshift (see their Fig.~2). The spectra of these LAEs do not exhibit any additional emission lines, preventing a definitive classification as either a star-forming galaxies or AGNs. These findings indicates that GS551 resides in a complex environment, compatible with the presence of a secondary AGN in GS551-B.

\subsection{Jekyll system (candidate dual AGN)}

The galaxy ZF-20115 (\citealt{SchreiberC2018}) is a post-starburst galaxy at $z\sim 3.7$ that exhibits a strong Balmer break and absorption lines. The far-infrared imaging reveals a luminous starburst located $\sim 0.4$\arcsec ($\sim 3$ kpc in projection) from the position of the rest-frame UV/optical emission, with an obscured star-formation rate of 100 \Msunyr (\citealt{Simpson2017}). The post-starburst and starburst galaxies were dubbed Jekyll and Hyde, respectively, by \cite{SchreiberC2018}.

NIRSpec IFS observations reveal the presence of additional companions in the surroundings of Jekyll and Hyde: streams and clumpy structures are observed in both rest-frame optical continuum and emission lines (Fig. \ref{fig:jekyllspec}). 
We identified two regions, on the north-east and west sides of Hyde, associated with AGN line ratios (Fig. \ref{fig:BPT}).  They are part of more extended structures dubbed Eastfield (hosting AGN-A) and Mr. West (hosting AGN-B), following \cite{Perez-Gonzalez2024}. A detailed investigation of the stellar population properties within this dense system is presented in \cite{Perez-Gonzalez2024}; here we report the stellar mass measurements for the two AGN host galaxies, of log(M$_*$) $= 9.0\pm 0.4$ (see Sect.~\ref{sec:stellarmasses}). 

In support of the presence of  dual AGNs, we also report  the presence of broad blue wings in the emission lines at the position of the two AGNs (up to $\sim 300-400$ \kms; see Fig. \ref{fig:jekyllspec}), possibly associated with outflows or, alternatively, to tidal interactions between the active galaxies and Jekyll \& Hyde. 

No X-ray emission is detected in the surroundings of this complex system; we therefore measured the upper limit on the 0.5--7 keV flux from the \chandra COSMOS-Legacy mosaic (\citealt{Marchesi2016}) as follows. We derive the net count rate by using as source region the circularised median of the \chandra PSFs at the position of Jekyll from each exposure ($\sim4.2$\arcsec) and a source-free circle of 15\arcsec-radius in the proximity of the target as background region. By assuming a photon index $\Gamma=1.8$, we obtain %considering a circular region with a radius of $\sim4.2$\arcsec (corresponding to the \chandra PSF) centred on Jekyll, 
$f_{0.3-7~\rm keV}=1\times10^{-16}$\ergs$\rm cm^{-2}$. Because of the size of the \chandra PSF ($\sim4.2$\arcsec) at the position of Jekyll, this upper limit refers to both AGNs we identified with NIRSpec.

\subsection{GS10578 system (triple AGN)}

GS10578 is a massive ($\sim 10^{11}$ M$_\odot$), post-starburst galaxy at $z\sim 3.06$, merging with lower-mass, gas-laden satellites (\citealt{DEugenio2024NatAs}); it also hosts an obscured AGN with a column density $N_{\rm H}=(5\pm4)\times 10^{23}$ cm$^{-2}$ and an X-ray luminosity $\log(L_{2-10\rm keV}/$\ergs) $=44.62\pm0.06$ (\citealt{liu2017_cdfs}). 
The inspection of archival MUSE observations allowed us to detect for the first time the extended \lya halo surrounding GS10578, as well as two additional \lya emitters (LAE1 and LAE2) at the same redshift of GS10578 (Fig. \ref{fig:lyaoiiimaps}); these structures are $\sim 50$ kpc (and $\sim 6000$ \kms) from another \lya halo discovered by \cite{Leclercq2017}.

The characterisation of the \lya halo surrounding GS10578 goes beyond the scope of this work. Here we focus on the study of the bright UV emission lines visible in the spectrum of GS10578, showing a double-peaked \lya line, as well as the \nv, \civ, \heii, and \ciii transitions, and the spectrum of the LAE2, with similar emission line features. The system LAE1, instead, is fainter and its spectrum only presents a \lya line at the same redshift of GS10578 and LAE2 (within a few hundreds of \kms).  
The spectra of GS10578 and LAE2 are  reported in Fig. \ref{fig:GS10578musespec}. 
From the modelling of the emission lines in LAE2 and GS10578, we inferred the flux ratios reported in Table \ref{tab:fit_MUSE}; these measurements are consistent with AGN ionisation, according to the UV diagnostics proposed by \cite[][Fig. \ref{fig:figUVBPT}]{Nakajima2022} and \cite{Feltre2016}. We note that the \ciii in LAE2 might be affected by atmospheric sky line residuals (see the 3$\sigma$ error curves associated with the LAE2 spectrum in the inset in Fig. \ref{fig:GS10578musespec}); however, the AGN classification is guaranteed using other UV diagnostics proposed by \cite{Feltre2016} and not involving the \ciii line, because of the \oiiiuv/\heii, \civ/\heii and other line ratios reported in Table \ref{tab:fit_MUSE}.

LAE1 and LAE2 have been also detected in JADES (\citealt{Rieke2023}), and identified with the ID 253554 and ID 197907, but associated with photometric redshift $z_{ph} =0.49 $ and 0.33, respectively. We used JADES photometry to determine the stellar mass of the \lya emitter hosting the AGN-C, log(M$_*$/M$_\odot$) $= 8.3 \pm 0.4$ (see Sect.~\ref{sec:stellarmasses}). 
LAE2 is not detected in 7 MS Chandra observations (\citealt{Luo2017}). Using the background map released by \cite{Luo2017}  
we derived an X-ray upper limit for LAE2 (AGN-C) as done for the Mr. West \& Eastfield AGN pair (median PSF: $\sim$2\arcsec; $\Gamma=1.8$): $f_{0.3-7~\rm keV}=9\times10^{-18}$ \ergs$\rm cm^{-2}$.

NIRSpec IFS observations cover a significant portion of the GS10578 \lya halo, but not LAE1 and LAE2 (see Fig. \ref{fig:lyaoiiimaps}). These observations allowed us to discover a few \oiii emitters in the surrounding of the primary AGN: the brightest one  hosts a secondary AGN (AGN-B in Fig. \ref{fig:lyaoiiimaps}), according to the BPT diagnostics. Being this source very close to the primary AGN ($\sim 4.7$ kpc), we  tested
whether the GS10578-B emission can be due to intrinsic or external ionisation source. We favoured the former scenario, as the primary AGN is not  powerful enough to ionise the gas at the position of GS10578-B (see Sect.~\ref{sec:ionisationsource}). Being this source very close to the primary AGN, it is likely that the X-ray luminosity inferred by \cite{liu2017_cdfs} may be contaminated by the AGN-B emission.

All in all, the presence of a secondary and a tertiary AGN  $\sim 0.7$\arcsec and $\sim 3.6$\arcsec of the GS10578 active nucleus, respectively, of the close LAE1 source, and a further relatively close ($\sim 50$ kpc) extended LAE make this system an excellent laboratory to study the early phases of a cluster in formation at $z \sim 3$.

\subsection{COS1656 (dual AGN)}

COS1656 is a massive ($\sim 10^{11}$ M$_\odot$) galaxy at $z\sim 3.5$.  
The NIRSpec IFS observations (Fig. \ref{fig:cos1656spec}) reveal two additional bright \oiii emitters at $\sim 10$ kpc north from the primary AGN. The spectra of the three targets are shown in Fig. \ref{fig:cos1656spec}: COS1656-A presents a strong outflow ($\sigma$(\oiii) $\sim 700$ \kms), while the other two sources show double-peaked emission lines. This is due to the partial overlap of the two \oiii emitters along our line of sight. According to the BPT diagnostics, the northern system is associated with a secondary AGN, with bright \oiii, \ha and \nii lines. This secondary AGN possibly contributes to the X-ray emission of COS1656 since the two sources are blended in the 7 MS CDFS data (\citealt{Luo2017}). As for the ionisation source of the close companion of AGN-B, it cannot be determined because of the undetected \nii.  

The secondary AGN is fairly well-detected in HST/ACS images, and associated with a stellar mass log(M$_*$/M$_\odot$)$ = 9.7\pm 0.3$   (see also \citealt{Weaver2022}). 
We note however that this measurement should be considered as an upper limit, because of the possible presence of two distinct systems on the LOS. 

The \chandra COSMOS Legacy observations do not allow for the identification of the dual AGNs, as their spatial separation is lower than the \chandra resolution at the position of the target (6\arcsec); the X-ray luminosity log($L_{2-10\ \rm keV}$/\ergs) $= 44.4$ reported by \cite{Marchesi2016} may therefore be attributed to both sources.

\section{Composite spectra}

\begin{figure*}[!htb]
\centering
\includegraphics[width=0.95\textwidth, trim=0mm 8mm 0mm 4mm]{{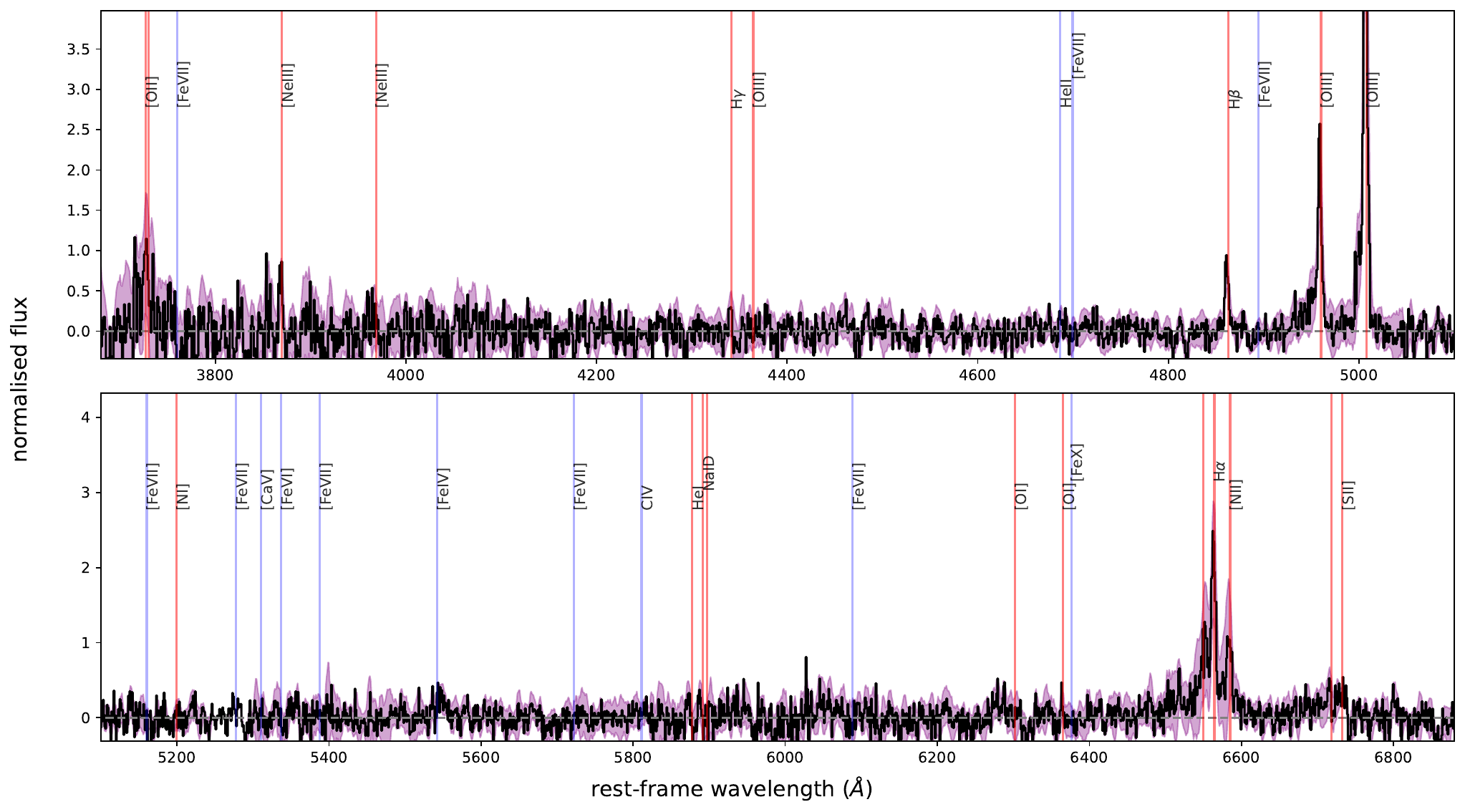}}
\caption{ Composite spectrum of the newly identified AGN at $z\sim 3$. The spectrum is obtained by combining the continuum-subtracted (0.15\arcsec $\times$ 0.15\arcsec) spectra of the newly discovered type 2 AGN COS1656-B, COS1638-B, GS551-B, GS10578-B, Eastfield, and Mr West, by employing a bootstrap method (\citealt{Perna2017b}); before the combine, all spectra were normalised to the \hb peak. Red vertical lines identify the main optical lines, while blue lines mark the position of coronal lines (CLs). No CLs are detected in the composite spectrum, but the most prominent emission lines show blue wings likely associated with outflows. }\label{fig:nonXraycomposite}
\end{figure*}

\begin{figure*}[!htb]
\centering
\includegraphics[width=0.95\textwidth, trim=0mm 8mm 0mm 4mm]{{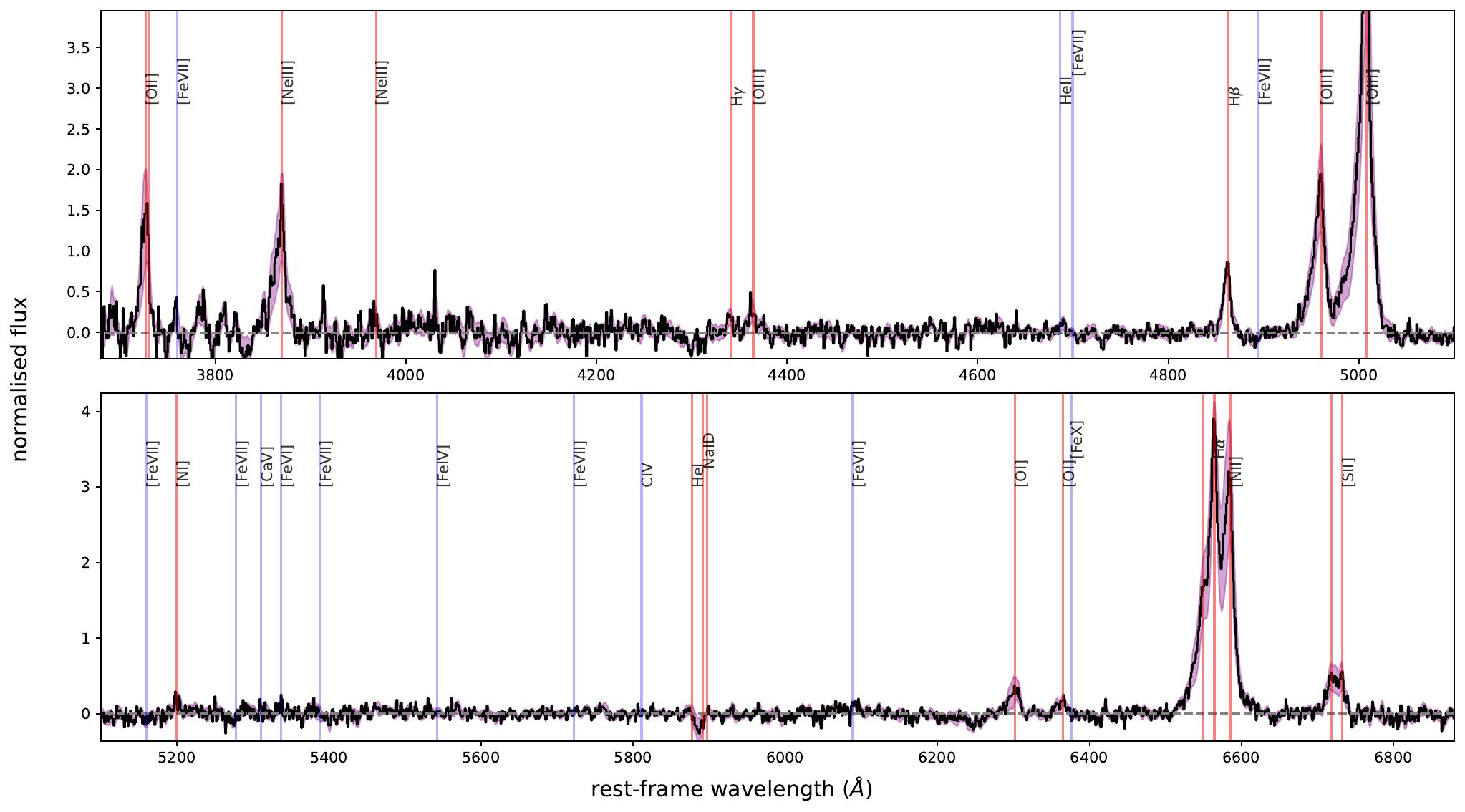}}
\caption{ Composite spectrum of X-ray detected AGNs at $z\sim 3$. The spectrum is obtained by combining the continuum-subtracted (0.15\arcsec $\times$ 0.15\arcsec) spectra of the X-ray detected type 2 AGNs GS10578-A, COS1656-A, GS551-A, GS133, GS19293, GS20936, by employing a bootstrap method (\citealt{Perna2017b}); before the combine, all spectra were normalised to the \hb peak. Red vertical lines identify the main optical lines, while blue lines mark the position of CLs. No CLs are detected in the composite spectrum. }\label{fig:Xraycomposite}
\end{figure*}

Figures \ref{fig:nonXraycomposite} and \ref{fig:Xraycomposite} show the composite spectrum of the newly identified type 2 AGNs at $z\sim 3$, and the composite spectrum of six X-ray detected AGNs at similar redshifts from GA-NIFS.

%\bigskip

%\bigskip

\section{Horizon-AGN dual-AGN fraction predictions}\label{sec:Ahorizon}

The predicted dual AGN fractions reported in Fig.~\ref{fig:AGNfraction} were calculated using the following general rules. The parent population includes all systems with at least one AGN having luminosity above a given threshold. Each system is counted as a maximum of one pair, even if it has multiple AGNs. Therefore, for a population of ten systems, if one has $\geq 1$ AGN companions, the dual AGN fraction is 1/10. This approach ensures a consistent calculation where each system is counted only once as a dual AGN, regardless of the number of active companions, and where the AGN population is a proxy for the number of massive galaxies undergoing galaxy mergers (e.g. \citealt{Chen2023, PuertoSanchez2024}).

We note that in Horizon-AGN, SMBHs are considered merged when their physical separation is less than 4~kpc. This could lead to an underestimation of the number of predicted dual and multiple AGN systems. Based on the distribution of physical distances in individual AGN systems, we estimated that the fractions in Fig. \ref{fig:AGNfraction} could be underestimated by about $\times5$\% for the sample with $d_{\text{p}} < 30$ kpc and $\times 10-20$\% for the sample with $d_{\text{p}} < 10$ kpc. Therefore, this effect should not be significant (see also Sect. 3.5 in \citealt{PuertoSanchez2024}). Additionally, we note that only one of our candidate dual AGN, GS551, has a projected separation of 2.9 kpc, which means its physical 3D separation could, in principle, be below 4 kpc.

\end{appendix}

%%%% End of aa.dem
\end{document}